\documentclass[journal]{IEEEtran}
\usepackage{amsmath,graphicx,mathtools,amsthm}
\usepackage[utf8]{inputenc}
\usepackage{epstopdf}
\usepackage{algorithm}
\usepackage{algorithmic}
\usepackage{amssymb}
\usepackage{extarrows}
\makeatletter
\def\fps@eqnfloat{t}
\def\ftype@eqnfloat{4}

\newenvironment{eqnfloat*}
{\@dblfloat{eqnfloat}} {\end@dblfloat} \makeatother
\usepackage{multirow}
\usepackage{float}
\usepackage{cite}
\usepackage{epsfig,latexsym}
\usepackage{breqn}
\usepackage{subcaption}
\usepackage{lipsum}
\usepackage{xcolor}
\usepackage[british]{babel}
\newtheorem{theorem}{Theorem}[]
\newtheorem{corollary}{Corollary}[theorem]

\def\sinc{\textrm{sinc}}

\def\AF{\textrm{AF}}
\def\BW{\textrm{BW}}
\def\SE{\textrm{SE}}
\usepackage{bm}
\usepackage{csquotes}
\usepackage{mathtools,amsthm}
\newtheoremstyle{case}{}{}{}{}{}{:}{ }{}
\theoremstyle{case}

\usepackage[justification=centering]{caption}

\usepackage{cite}
\usepackage{epsfig,latexsym}
\usepackage{amsmath}
\def\sinc{\textrm{sinc}}

\title{Frequency Diverse Array Radar: New Results and Discrete Fourier Transform Based Beampattern}
\author{Muhammad Zubair, Sajid Ahmed {\em Senior Member IEEE},  and Mohamed-Slim Alouini {\em Fellow IEEE}}

\begin{document}

\maketitle
\begin{abstract}
In the phased-array radar (PAR) signals from each antenna are transmitted at the same carrier frequency, which yields narrowly focused only angle dependent beampattern. In contrast, in the frequency-diverse-array (FDA) radar signals from antenna array are generally transmitted at linearly increasing frequencies that yields range, time, and angle dependent beampattern.  
Reported literature on FDA radar missed the contribution of path-differences in the signal model due to the antenna array elements, which may lead to misleading results. In this work, incorporating missed path-differences, the signal model of FDA radar is corrected. Using the corrected signal model, it is shown that the instantaneous beampattern depends on the number of transmit antenna and average beampattern depends on the product of frequency-offset and pulse-duration.  
Moreover, to illuminate the desired region-of-interest for longer dwell time, discrete-Fourier-transform based low-complexity algorithm is proposed. In contrast to the conventional FDA radar's  \enquote{S} shaped beampattern, the beampattern of the proposed algorithm changes linearly with range. Simulation results compare the performance of our proposed algorithm with the existing ones and show the superiority of our proposed algorithm. 
\end{abstract}  

\section{Introduction}\label{sec:Intro}
{\huge P}hased-array radars (PARs) have aspiring features of adaptive and agile electronic beam steering, due to which they are replacing conventional mechanically beam steering radars \cite{phasedpresentpastfuture,phaseddevelopment,phasedhandbook}. The initial use of PAR was restricted to the military applications only but over the time, they are being used in numerous civilian applications, such as radar-based land mobile communication, air traffic monitoring, bio-medical, and adaptive cruise control \cite{phasedreview,phasedbenifits,phasedautomotive1,phasedautomotive2,phasedmedicine}. The benefits of PARs come at the cost of multiple debase factors, such as their complex and costly structure, scan loss due to the low dynamic range of phase-shifters, and only angle-dependent beampattern \cite{phasedhandbook,modernradarprinciples}. These debase factors limit the deployment of PARs in many other high tech commercial applications \cite{phasedthesis1}.

Recently, a new framework called frequency-diverse-array (FDA) radar is proposed \cite{4antonik2006range}. In conventional  FDA radar, linearly increasing frequency-offsets (FOs) are applied across the antenna array elements that result in periodic range, angle, and time-dependent beampattern \cite{secmen2007frequency}. The FDA radar beampattern in multiple dimensions can potentially be exploited to address the challenges of cost, complexity, and scan loss in the PAR. To steer the beam, in contrast to the use of costly and low-dynamic range phase-shifters in PAR, FOs can be optimized in FDA radar. Alternatively, the transmitted signal from each antenna can be multiplied by appropriate weights to steer the beam in FDA radar \cite{antonikthesis}, these weights can also be exploited to change the shape of beampattern. The FDA radar range dependent beampattern can suppress known interferers at different ranges and address the issues of scan loss \cite{4antonik2006range,wang2016overview}.

Conventional FDA radar yields time dependent periodic \enquote{S} shaped narrow beam with respect to range and angle. To illuminate an object, the power should be focused on it for longer dwell times. However, due to the time-dependent beam, the focus of the FDA radar beam changes continuously, which results in shorter dwell time. Due to which the FDA radar can miss the weak targets with high probability.  
To resolve the time-dependent beam issue a pulsed FDA radar scheme is proposed in \cite{pulsed_FDA}. In this scheme, the achievement of quasi-stationary beampattern is claimed by transmitting a very short duration pulse without considering the influence of the time variable. To steer the beam in different directions, the proposed scheme suggests the application of corresponding FOs across the antenna array. The fundamental drawback of this scheme is its short dwell time. To increase the dwell time at the given range, in \cite{TimeInvariant1khan} time-dependent FOs are applied across the antenna array. In this scheme, although a pulse is applied across the antenna array with time-dependent FOs, the beampattern is periodic due to which signal to noise-plus-interference ratio may decrease. The advantage of this scheme is that multiple targets can be tracked. To increase the dwell time at the given range, in \cite{TimeInvariant2shao} weights of the FDA radar are optimized using CVX toolbox \cite{CVX} and updated after each pulse. Due to which the computational complexity of this scheme is quite high. 
For aperiodic beam, a non-uniform inter-element spacing FDA radar is proposed in \cite{nonuniformFDA}. Since this strategy is dependent on the inter-element spacing between the antenna elements that is not possible to change in real-time, this scheme is not feasible for adaptive applications. Moreover, the development of such a system requires much higher precision compared to the uniform-linear array. 
To achieve aperiodic and time-invariant beam within a pulse to illuminate single target, a scheme is proposed in \cite{Timeinvariant3yao}. While to illuminate multiple targets another scheme is proposed in \cite{Timeinvariant5yao}. In these schemes, time modulated logarithmically increasing FOs are applied across the antenna array. 

In \cite{Timeinvariant6chen,Timeinvariant7chen}, it is pointed out that most of the FDA radar algorithms mentioned above ignore the propagation delay\footnote{A time required by a signal to propagate from the transmitting antenna to the target.} and show beampattern at $t=0$ or $t=T$. However, for a target at range $R_o$, when $t<t_o=\frac{R_o}{c}$ none of the transmitted signal reach the target, where $t_o$ is the propagation delay and $c$ is the velocity of light. Similarly, if $T < t_o$ none of the transmitted signal reach the target at $t=T$. Therefore, in both cases, the beampattern will be zero. The work in \cite{Timeinvariant6chen,Timeinvariant7chen} modify the FDA signal model by incorporating the propagation delay $t_o$. Even their modified signal model ignores delays due to the path-differences of antenna array elements that give rise to transient and steady-state beampatterns as discussed in Sec. \ref{Sec:SigMod}.

To address the above-mentioned issues, in this paper 
\begin{itemize}
    \item Signal model of FDA radar is corrected by incorporating the ignored time delays due to the path-differences of the uniform-linear-array (ULA) elements.
    \item A mathematical relationship between the beamwidth of instantaneous beampattern and number of antenna elements is developed. 
    \item Another mathematical relationship between the average power received by a target at a given range and the product $f_oT$ is developed and its significance is discussed. 
    \item A bound is devised for the value of $f_oT$ that must be not violated while selecting their values for radar system under operation.
    \item To illuminate the desired spatial region at a given range, the weights of FDA radars are derived in closed-form by exploiting discrete-Fourier-transform (DFT), which has much lower computational complexity compared to the reported algorithms \cite{TimeInvariant2shao,wang2013range} and allow the control of dwell time.
\end{itemize}
 
The remainder of the paper is organized as follows: The FDA radar signal model is discussed in Sec. \ref{Sec:SigMod}, Sec. \ref{Sec:Novel_Derivations} derives some novel results for FDA radar that are not reported yet in the literature. To focus the transmitted power in the desired spatial region for the given range the algorithm is proposed in Sec. \ref{Sec:PrpsdSchm}, simulation results are given in Sec. \ref{Sec:Simulations}, and conclusions are drawn in Sec. \ref{sec:con}.

\textbf{Notations:} Bold upper case letters, ${\bf X}$, denote matrices while lower case letters, ${\bf x}$, denote vectors. Transpose and conjugate transposition of a matrix are respectively denoted by $(\cdot)^T$ and $(\cdot)^H$. The conjucgate of a scalar is denoted by $(\cdot)^*$. The close interval $\{x: a \le x \le b\}$ is denoted as $[a,b]$.  

\section{Signal Model} \label{Sec:SigMod}
\begin{figure}
    \centering
    \includegraphics[height=3in,width=3.5in,keepaspectratio]{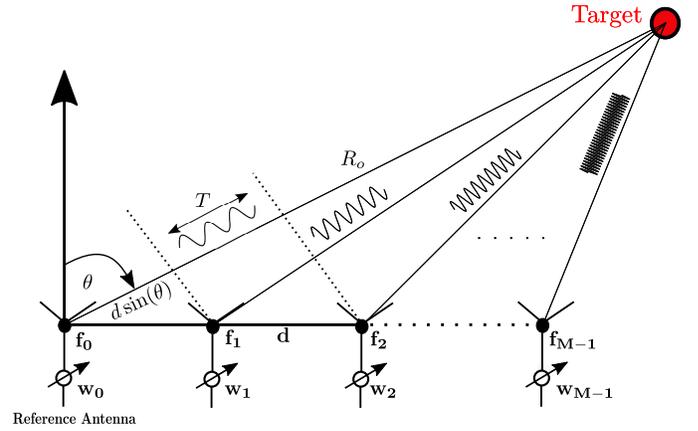}
    \caption{Baseband model of a conventional FDA radar transmitter.}
    \label{fig1:ULA_FDA}
\end{figure}

The baseband model of an FDA radar is shown in Fig. \ref{fig1:ULA_FDA}. The number of antenna in the array is $M$, the distance between any two adjacent antennas is $d$, the transmitted signal from the $m$th antenna is multiplied by a weight $w_m$, and the FO introduced in the carrier-frequency of the $m$th antenna is $f_m$. With these parameters, the transmitted signal from the $m$th antenna can be written as 
\begin{align}
s_{m}(t) &=  w_{m} e^{-j2\pi (f_c + f_m)t}, \quad ~\!\quad t \in \left[0, T\right]\notag \\
         & = 0,  \quad\qquad\qquad\qquad \qquad t<0~ \mbox{or}~t>T
\end{align}
where $m=0,1,\ldots,M-1$. If a target is present in the far field at a distance $R_o$ from the reference antenna in the direction of $\theta$, the transmitted signal from the $m${th} antenna will cover a distance $R_o - md\sin(\theta)$ to reach target. If $f_m=mf_o$, the superposition of $M$ transmitted signals at the target can be written as
\begin{align}
\label{eq3:rt_1}
    r(t) &= \sum_{m=0}^{M-1} w_{m}  e^{-j2\pi (f_c + mf_o)(t-\frac{R_o-md\sin(\theta)}{c})}.
\end{align}
Assuming $t_o = \frac{R_o}{c}$ and $\tau_{m}(\theta) = \frac{md\sin(\theta)}{c}$, (\ref{eq3:rt_1}) can be written as
\begin{align}
   \label{eq3:rt_2}
    r(t)  &= \sum_{m=0}^{M-1} w_{m}  e^{-j2\pi (f_c + mf_o)(t-t_o +\tau_{m}(\theta))}.  
\end{align}
%
%
\begin{figure*}
    \centering
    \includegraphics[height=3.5in,width=7in,keepaspectratio]{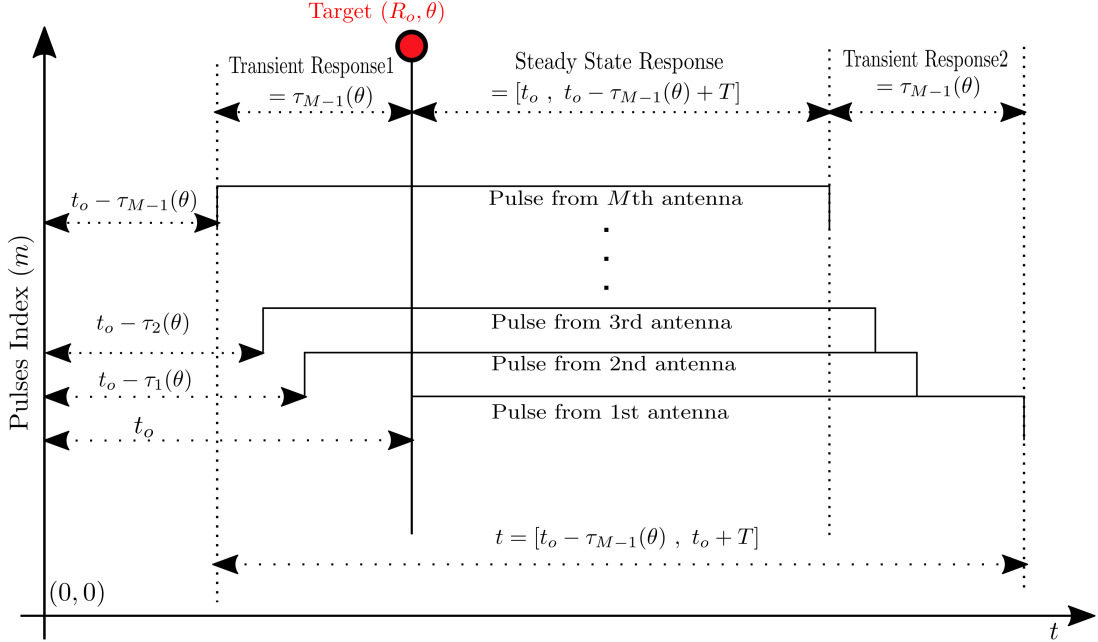}
    \caption{Timing diagram of the propagation of transmitted signals from different antennas for the target at range $R_o$ and angular location $\theta$.}
    \label{fig2:accurate_time}
\end{figure*}
It can be observed that the start of the transmitted signal from the $m$th antenna will be received at the target after $\frac{R_o-md\sin(\theta)}{c}$ seconds. Therefore, during the time interval $t_o -\tau_{M-1}(\theta)$ to $t_o -\tau_{M-2}(\theta)$ the target will be illuminated only by the $M$th antenna, while during the time interval $t_o -\tau_{M-2}(\theta)$ to $t_o -\tau_{M-3}(\theta)$ the target will be illuminated by the $M$th and $(M-1)$th antennas, and so on. This way after $t_o$ the target will be illuminated by all antennas. The beampattern between the time interval $t_o -\tau_{M-1}(\theta)$ to $t_o$ is called a transient beampattern. After time $t_o$ all the transmitted signals will be contributing to illuminate the target at the range $R_o$, therefore the corresponding beampattern can be called steady state beampattern. Since all the transmitted signals have a pulse duration of $T$, the contribution of each transmitted signal in the received signal will be only for $T$ seconds. Due to which, the contribution of $M$th transmitted signal will be lost after $t_o -\tau_{M-1}(\theta) + T$ seconds, similarly, the contribution of $(M-1)$th transmitted signal will be lost  after $t_o -\tau_{M-2}(\theta) + T$ seconds, and so on. Observing this, it can be said that after time $t_o-\tau_{M-1}(\theta)+T$ seconds, the contribution of the transmitted signals to illuminate the target at the range $R_o$ and angle $\theta$ will start decreasing. Again after $t_o-\tau_{M-1}(\theta)+T$ seconds, the corresponding beampattern will be called as a transient beampattern. Arrival times of pulses from different transmit antennas at the target are shown in Fig. \ref{fig2:accurate_time}. From the above discussion, it can be concluded that target with coordinates $R_o$ and $\theta$ will be illuminated with different number of antennas in the first, second, and  third time intervals. Therefore, in the selected processing time interval all the transmitted signals should be illuminating the target. 
Now, it is obvious that before $t_o-\tau_{M-1}(\theta)$ target with coordinates $R_o$ and $\theta$ cannot be illuminated. This fact is ignored in most of the available literature, for example  \cite{Timeinvariant6ref1,Timeinvariant6ref3,Timeinvariant6ref4,Timeinvariant6ref5,Timeinvariant6ref6,tiszeroma2019general}, where the illumination of target with respect to range and angle is shown at $t=0$, which is not possible. The illumination time should be started at-least at $t=t_o-\tau_{M-1}(\theta)$ so that at-least one transmitted signal is arrived at the target. However, the actual illumination time should start at $t=t_o$ so that all the antennas can contribute in the illumination of target.
Similarly, the research presented in \cite{TimeInvariant1khan,TimeInvariant2shao,Timeinvariant3yao,Timeinvariant4yao,Timeinvariant5yao,Timeinvariant6liao} shows the target illumination with respect to range and angle between $t=0$ and $t=T$, which is again misleading. Similar to $t=0$, the choice of $t=T$ is incorrect because if $t_o$ is greater than $T$, the target will not illuminate. This fact is also pointed out in \cite{Timeinvariant6chen,Timeinvariant7chen}, they address this issue by showing the illumination of target with respect to range and angle between $t=t_o$ and $t=t_o +T$. Even in this work, the first and second transient interval are ignored. Moreover, the precise information of the time interval in which all transmit antennas contribute to illuminate the target is not discussed at all. After $t=t_o-\tau_{M-1}(\theta)+T$, the illumination of the target will follow the transient-2 state. Therefore, to have reliable detection of the target, it is very important to carefully select the target illumination times.

To find the array-factor (AF) of an FDA radar, rearranging the terms in (\ref{eq3:rt_2}), the received signal can be written as   
\begin{align}
   r(t) &= e^{-j2\pi f_c (t-\frac{R_o}{c})}\sum_{m=0}^{M-1} w_{m} e^{-j2\pi \Phi_{m}},
    \label{eq3:rt_3}
\end{align}
where $\Phi_{m} = m\left(f_o(t-\frac{R_o}{c}) +f_c \frac{d \sin(\theta)}{c} +\frac{m f_o d\sin{\theta}}{c} \right)$. In (\ref{eq3:rt_3}), by separating the terms depending on the geometry of the radar system, the AF can be defined as
\begin{eqnarray}
\AF (t,R_o,\theta) \!\!\!\!&=& \!\!\!\!\! \!\sum_{m=0}^{M-1} w_{m} e^{-j2\pi \Phi_{m}}, \mbox{~for }t \in \left[t_o-\tau_M(\theta), t_o+T\right]  \notag,   \\  
&=& 0,  \quad   t_o-\tau_M(\theta) > t > t_o+T, 
\label{eq:AF1}
\end{eqnarray}
which can be used to find the beampattern of the FDA radar with respect to $t$, $R_o$, and $\theta$ as
\begin{align}
    B (t; R_o,\theta) &= |\AF(t,R_o,\theta)|^{2}.
    \label{eq:BP}
\end{align}
For the given time, the states of the beampattern are defined in Table \ref{Table:1}.
\begin{figure*}
    \centering
    \includegraphics[height=3.5in,width=7in,keepaspectratio]{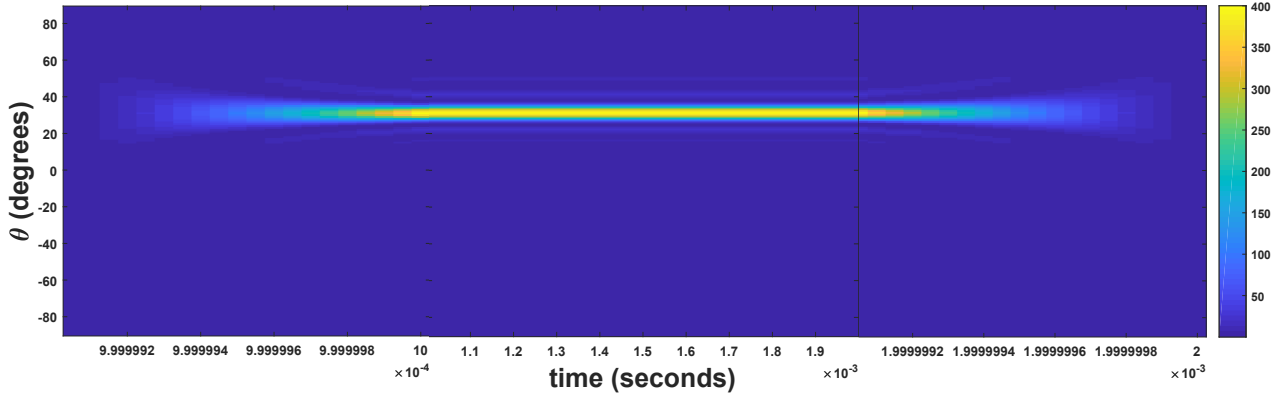}
    \caption{Transient and steady state beampattern of an FDA radar at range $R_o$.}
    \label{fig2:Simu_accurate_time}
\end{figure*}
%
\begin{table}[]
\caption{States of beampattern Response}
    \centering
   {\large
\begin{tabular}{|c|c|}
\hline
$t_o-\tau_{M-1}(\theta) \rightarrow t_o$     &  {\em Transient-1 state} \\
\hline
$t_o \rightarrow t_o-\tau_{M-1}(\theta)+T $     & {\em Steady State} \\
\hline
$ t_o-\tau_{M-1}(\theta)+T \rightarrow t_o +T$     & {\em Transient-2 state} \\ 
\hline
\end{tabular}
}
 \label{Table:1}
\end{table}
To show the effect of actual time on the beampattern, a simulation is performed for a target located at $R_o=300$ km and $\theta=30^o$. Radar parameters considered for this simulation are $M=20, f_c=5$ GHz, $f_o=100$ Hz and the corresponding beampattern is shown in Fig. \ref{fig2:Simu_accurate_time}.  In the figure, it can be observed, the beampattern is changing in the transient state-1. The reason is obvious, as discussed earlier in the transient state-1, the contribution of signals from the transmit antennas is increasing with time. Initially, the beampattern is due to a single antenna element so it should be isotropic. Therefore, in the beginning at $t=t_o-\tau_{M-1}(\theta)$, the beampattern is quite wide. Then after each $\tau(\theta)$ seconds, the contribution of signal from the adjacent transmit antenna in the illumination of target is added until $t=t_o$ seconds. After the addition of signals from all $M$ antennas, the beampattern is stabilized and not changing further. This state is called a steady state of beampattern. After time $t_{o}-\tau_{M-1}(\theta)+T$ the contribution of the signals in the illumination start decreasing and opposite to state-1 beampattern can be observed.  Time constraint steady state beampattern can be defined as
{\small
\begin{equation}
B_{ssr}(t;R_o,\theta) = \left|\sum\limits_{m=0}^{M-1} w_{m}e^{-j2\pi \Phi_{m}}\right|^2 \!, t \in \left[t_{o}, t_{o} - \tau_{M-1}(\theta) + T\right],\label{eq:BP_ssr}
\end{equation}}
where the subscript \enquote{$ssr$} in $B_{ssr}$ is used for steady state response. 
%
\section{Some Novel Derivations of FDA Radars}\label{Sec:Novel_Derivations}
In this section some new results for FDA radar are derived that are not reported yet in the literature. 
%
\subsection{Beamwidth of Conventional FDA Radars} 
\begin{figure}[htbp]
    \centerline{\psfig{figure=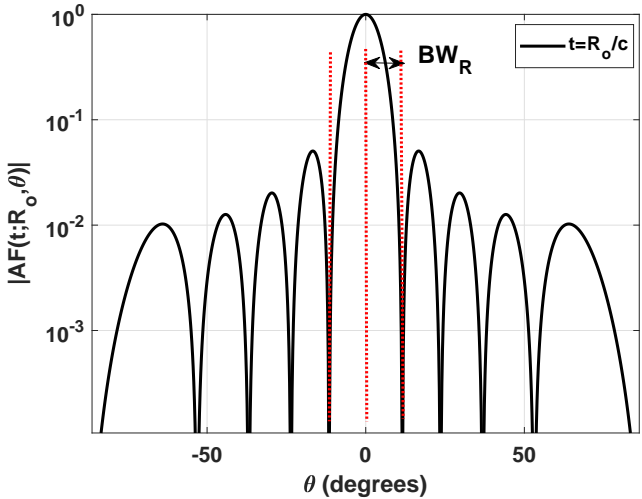,width=3.7in} }
    \caption{Beamwidth of FDA radar for  $\phi_o=0$, $M=10$, and $f_o=100$ Hz. For $f_o\left(t-\frac{R_o}{c}\right)>0$ the beampattern will shift left.}
    \label{fig:Beamwidth_FDA}
\end{figure}

{\label{subSec:BeamwidthDerivation}}
In FDA radar incremental frequency $f_o$ is kept much lower than the carrier frequency $f_c$. As a consequence, the term $\frac{m^2 f_o d \sin{\theta}}{c} \ll  \frac{m f_c d \sin{\theta}}{c}$ and it can be ignored in (\ref{eq:AF1}) to write AF in more compact form as
\begin{align}
\AF (t,R_o,\theta) &= \sum_{m=0}^{M-1} w_{m} e^{-j2\pi [mf_o(t-\frac{R_o}{c}) +m\frac{ \sin(\theta)}{2}) ]}
 \label{eq:AF_1}.
\end{align}
Assuming $w_m=e^{-jm\phi_o}$, AF can be written as 
\begin{align}
\AF(t,R_o,\theta) &= \sum_{m=0}^{M-1} e^{-j2\pi m \left[f_o(t-\frac{R_o}{c}) + \frac{\phi_o}{2\pi}+ \frac{ \sin(\theta)}{2}) \right]}. \label{eq:AF_2}
\end{align}
Applying sum of geometric series formula on (\ref{eq:AF_2}), we can write
\begin{align}
\AF(t,R_o,\theta)  &=\frac{1-e^{-j2\pi M\left[f_o(t-\frac{R_o}{c}) + \frac{\phi_o}{2\pi} + \frac{ \sin(\theta)}{2}) \right]}}{1-e^{-j2\pi\left[f_o(t-\frac{R_o}{c}) + \frac{\phi_o}{2\pi} + \frac{ \sin(\theta)}{2}) \right]}}, \notag\\
  &= e^{-j\pi (M-1) \left[f_o(t-\frac{R_o}{c}) + \frac{\phi_o}{2\pi} + \frac{ \sin(\theta)}{2}) \right]}\notag\\
  &\times \frac{\sin({\pi M \left[f_o(t-\frac{R_o}{c}) + \frac{\phi_o}{2\pi} + \frac{ \sin(\theta)}{2}) \right]})}{\sin\left({\pi\left[f_o(t-\frac{R_o}{c}) + \frac{\phi_o}{2\pi} + \frac{ \sin(\theta)}{2}) \right]}\right)}
  \label{eq:AF3}.
\end{align}
%
%
%
Using (\ref{eq:AF3}), the absolute value of AF can be written as
\begin{equation}
|\AF(t,R_o,\theta)| =  
 \frac{\sin\left({\pi M \left[f_o(t-\frac{R_o}{c}) + \frac{\phi_o}{2\pi} + \frac{ \sin(\theta)}{2}) \right]}\right)}{\sin\left({\pi\left[f_o(t-\frac{R_o}{c}) + \frac{\phi_o}{2\pi} + \frac{ \sin(\theta)}{2})\right]}\right)},
  \label{eq:AF6}
\end{equation}
which is a $\sinc(\cdot)$ function. It is well know that the maximum value of this function is $M$ \cite{Book:SNS} and occurs at $\psi(t,\theta)= 0$ and the nulls of this function will occur at $\psi(t,\theta) = n\pi$ for $n=\pm1,\pm 2, \ldots$ and so on. Let us define the beamwidth of array factor as the Rayleigh beamwidth (distance from peak to first null). The first null will occur whenever
\begin{align}
    M\pi\left[f_o\left(t-\frac{R_o}{c}\right) + \frac{\phi_o}{2\pi} + \frac{ \sin(\theta)}{2}\right] = \pi. \notag
\end{align}
At $t=\frac{R_o}{c}$, the first null in the value of array factor will occur when $\frac{M\sin(\theta)}{2} + \frac{\phi_o}{2\pi} = 1$. From this the angular location of first null can be derived as  
\begin{align}
   \theta_{1N} = \sin^{-1}\left(\frac{2}{M}-\frac{\phi_o}{\pi}\right). 
\end{align}
Similarly at $t = \frac{R_o}{c}$, the maximum value of AF will occur when $\frac{\sin(\theta)}{2} + \frac{\phi_o}{2\pi} = 0$, which can be used to derive the angular location of maximum value of AF as   
\begin{align}
    \theta_{\max} &= \sin^{-1}\left(-\frac{\phi_o}{\pi}\right) 
    \label{eq:Peak}.
\end{align}
Therefore, the Rayleigh beamwidth of conventional FDA radar system can be calculated as
\begin{align}
    \BW_{R} &= \theta_{1N} - \theta_{\max} \notag\\
    &= \sin^{-1}\left(\frac{2}{M}-\frac{\phi_o}{\pi}\right)  + \sin^{-1}\left(\frac{\phi_o}{\pi}\right)
    \label{eq:BW_R}.
\end{align}
By using Maclaurian series approximation of $\sin^{-1}(\cdot)$ function given by $\sin^{-1}(x) = x + \frac{x^3}{6} + \frac{3x^5}{40} + \ldots$, in (\ref{eq:BW_R}), the approximate value of Rayleigh beamwidth can be calculated as

\begin{align}
   \BW_{R} \approx \frac{2}{M} + \frac{\phi_o^{2}}{M {\pi}^2}. 
     \label{eq:BW_R1}
\end{align}
Note that in (\ref{eq:BW_R1}) the Maclaurian series terms of  power five and more are discarded due to their negligible values compared to the selected terms. Expression (\ref{eq:BW_R1}) shows that at $t=\frac{R_o}{c}$ the Rayleigh beamwidth mainly depends on the number of antennas and slightly on the initial phase while it does not depend on $f_o$ or the duration of pulse. Three corollaries can be obtained from the AF derived in (\ref{eq:AF6}) and are given below
\begin{corollary}
If $f_o = 0$, the argument of $\sin(\cdot)$ function in the numerator $\pi M\left[f_o\left(t-\frac{R_o}{c}\right) + \frac{\phi_o}{2\pi} + \frac{ \sin(\theta)}{2}\right]$ and in the denominator $\pi \left[f_o\left(t-\frac{R_o}{c}\right) + \frac{\phi_o}{2\pi} + \frac{ \sin(\theta)}{2}\right]$ becomes independent of range $R_o$ and time $t$. Therefore, $f_o=0$ will result in traditional phased-array beampattern.  
\end{corollary}
\begin{corollary}
At $t = \frac{R_o}{c}$ and $f_o\ne 0$, the argument of $\sin(\cdot)$ function in the numerator $\pi M\left[f_o\left(t-\frac{R_o}{c}\right) + \frac{\phi_o}{2\pi} + \frac{ \sin(\theta)}{2}\right]$ and in the denominator $\pi \left[f_o\left(t-\frac{R_o}{c}\right) + \frac{\phi_o}{2\pi} + \frac{ \sin(\theta)}{2}\right]$ becomes independent of range $R_o$. At this instant the beampattern will follow the beampattern of phased-array radar with a boresight shifted through an angle $\frac{\phi_o}{2}$ as shown in Fig. \ref{fig:Beamwidth_FDA}.
\end{corollary}
\begin{corollary}
For any value of $\frac{R_o}{c}-\tau_{M-1}(\theta) < t \leq T+\frac{R_o}{c}-\tau_{M-1}(\theta)$, the beampattern will again follow the phased-array radar beampattern with a boresight shifted through an angle $\left(f_o\left(t-\frac{R_o}{c}\right) + \frac{\phi_o}{2\pi} \right)$.  
\end{corollary}

Since, at $\frac{R_o}{c}-\tau_{M-1}(\theta) < t \leq T+\frac{R_o}{c}-\tau_{M-1}(\theta)$, the term $f_o\left(t-\frac{R_o}{c}\right)$ and $\phi_o$ do not depend on the value of $\theta$, however, these terms will spatially move the AF pattern through an angle $2f_o(t-t_o) + 2\frac{\phi_o}{\pi}$ ($\phi_o>0$) radians on the left hand side, this phenomena can be called as spatial-exploration ($\SE$) of FDA radar system. 
\subsection{Average Transmit Beampattern of FDA Radar} \label{Sec:AvgBeamPattern0}
\begin{figure}
    \centering
    \captionsetup{justification=centering}
    \includegraphics[scale=0.6]{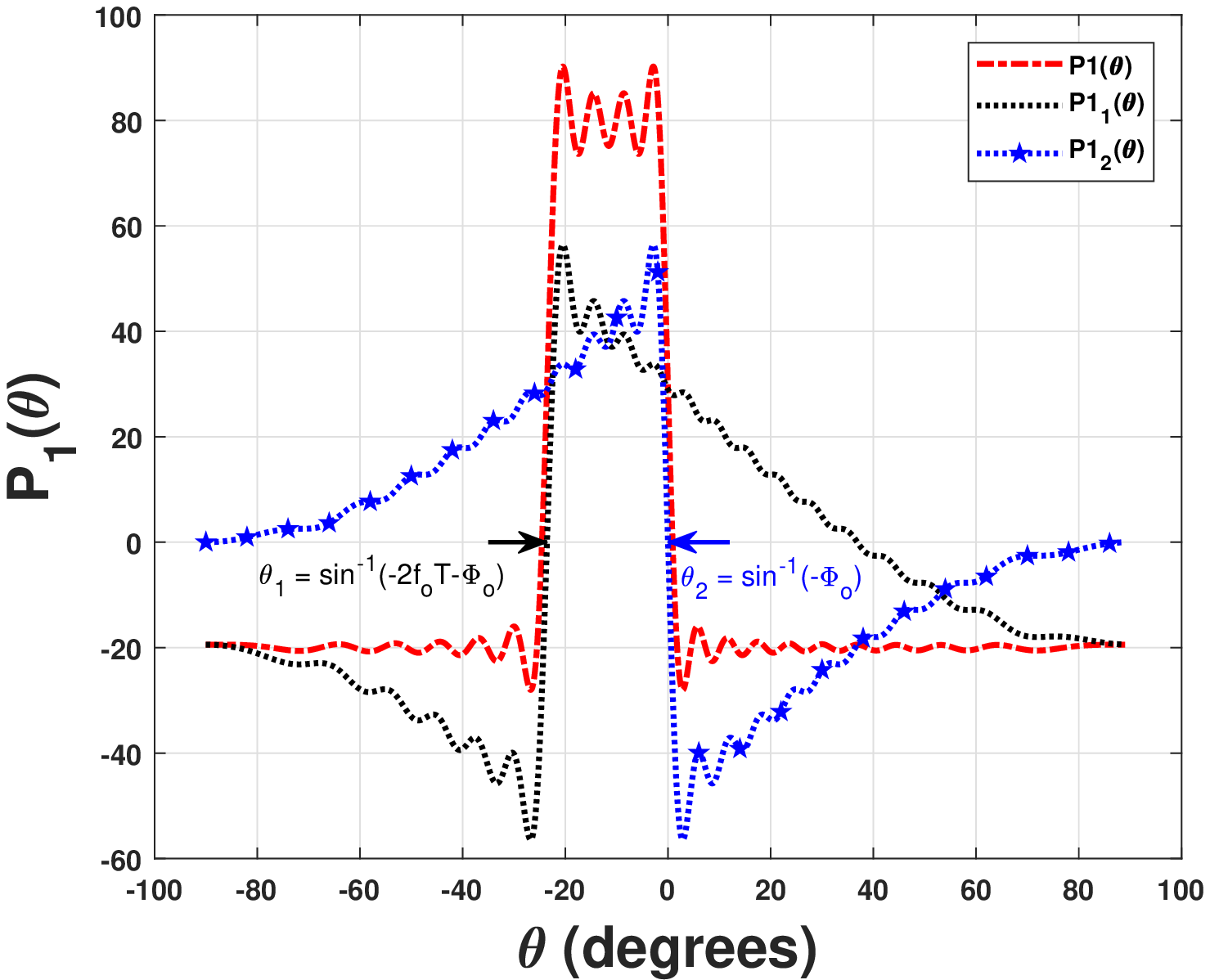}
    \caption{Average beampattern of $P1(\theta)$, $P1_1(\theta)$, $P1_2(\theta)$. Initial phase $\phi_o=0^o$, $M=20$,$f_o=200$Hz,$T=1m$s.}
    \label{fig:BW_Avg_Power1}
\end{figure}

\begin{figure}
    \centering
    \captionsetup{justification=centering}
    \includegraphics[scale=0.6]{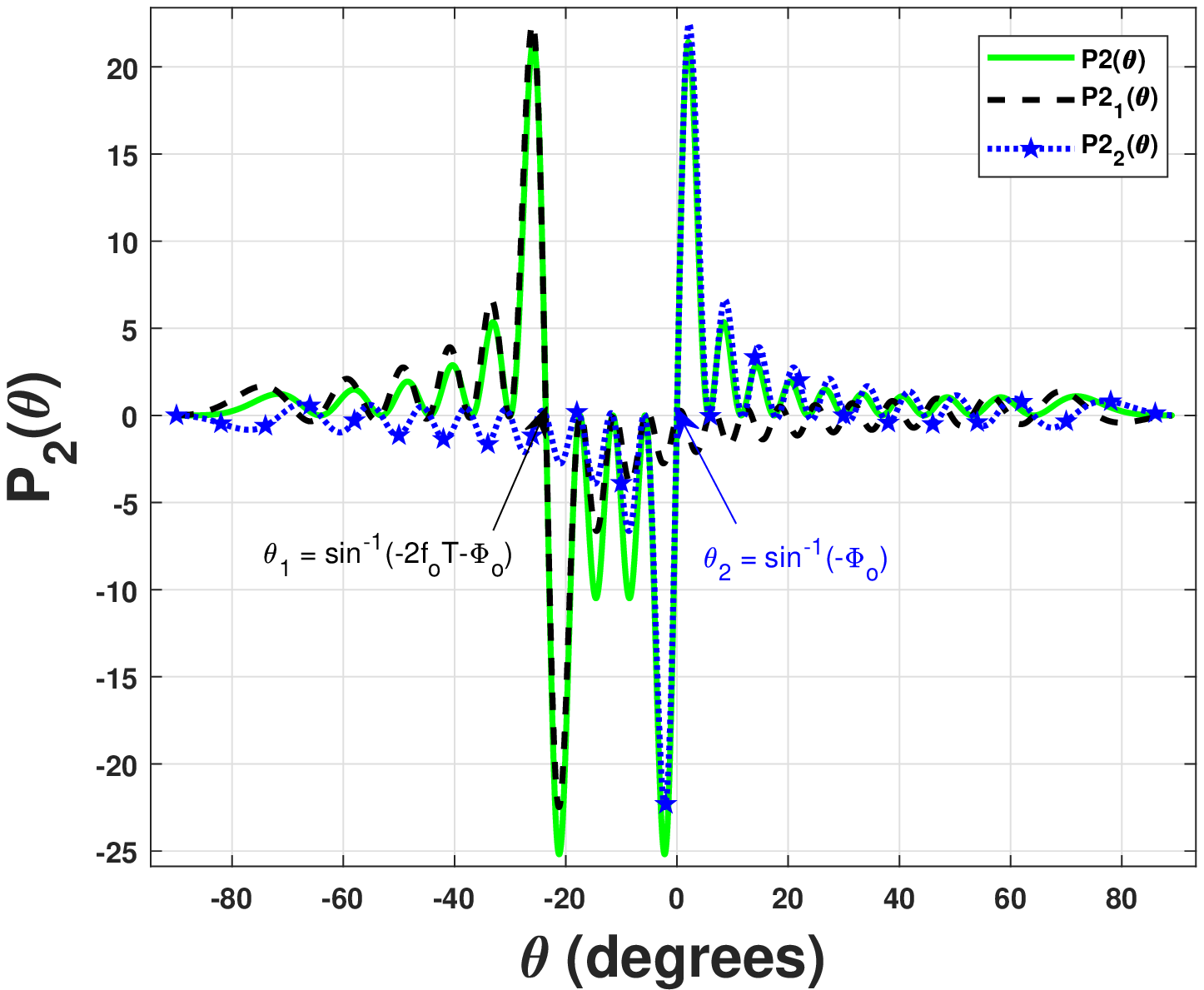}
    \caption{Average beampattern of $P2(\theta)$, $P2_1(\theta)$, $P2_2(\theta)$. Initial phase $\phi_o=0^o$, $M=20$,$f_o=200$Hz, $T=1m$s.}
    \label{fig:BW_Avg_Power2}
\end{figure}

\begin{figure}
\centering
\captionsetup{justification=centering}
\includegraphics[scale=0.6]{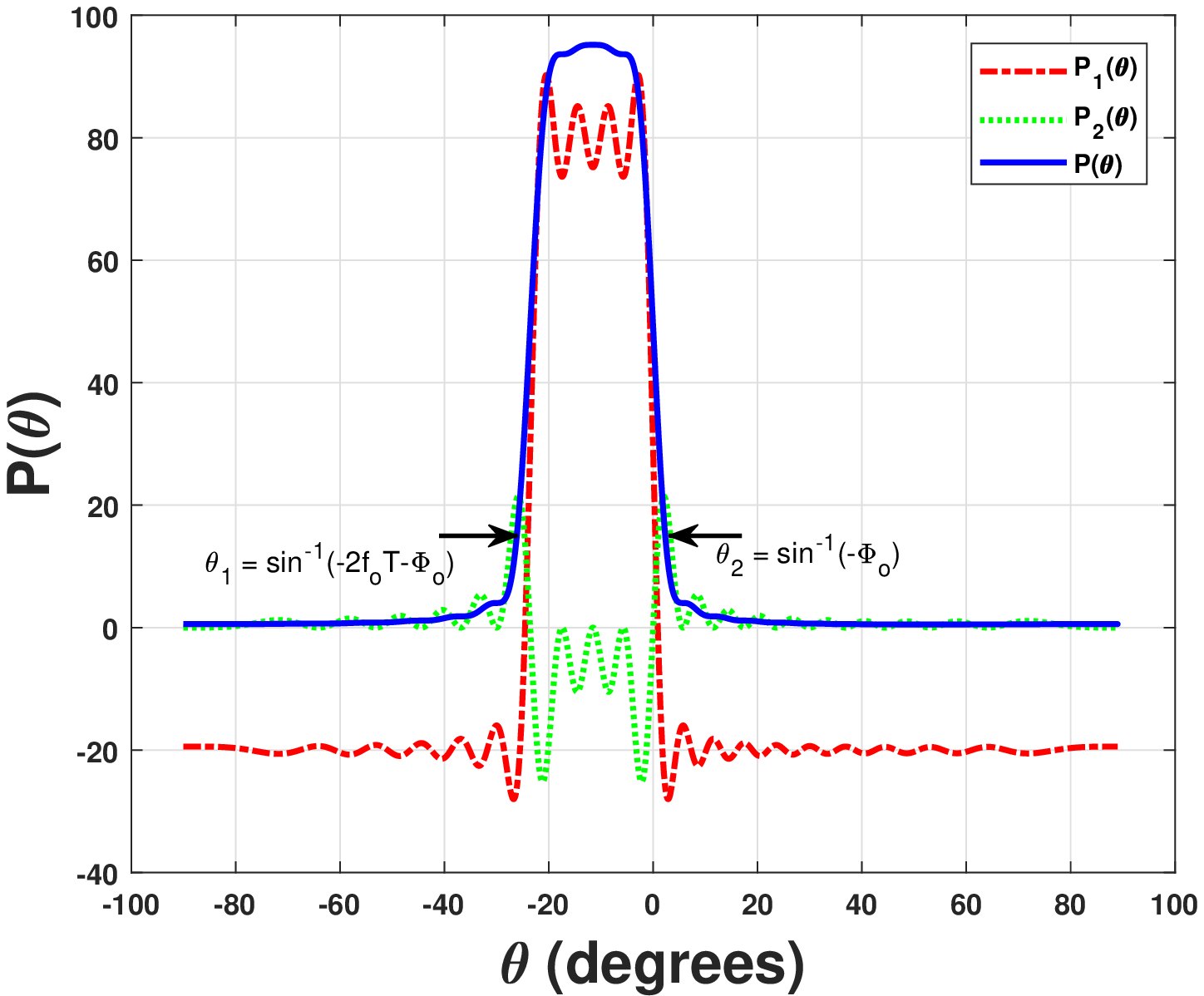}
\caption{Average beampattern of $P(\theta)$, $P1(\theta)$, and $P2(\theta)$. Initial phase $\phi_o=0^o$, $M=20$, $f_o=2k$Hz, $T=1m$s.}
\label{fig:BW_Avg_Power}
\end{figure}

The theoretical value of SE intuitively is derived in \cite{Timeinvariant6chen}, which lacks proper mathematical reasoning. In this section, an expression for the average beampattern is derived that helps to compute the SE of FDA radar (also called angle spread of beampattern). 
To derive the average transmit beampattern of FDA radar, again by exploiting the fact $\frac{m^2 f_o d \sin{\theta}}{c} \ll  \frac{m f_c d \sin{\theta}}{c}$ in (\ref{eq3:rt_3}), we can write 
\begin{equation}
r(t,R_o,\theta) = p(t,R_o)\sum_{m=0}^{M-1} w_m
e^{-j2\pi \left[f_o(t-\frac{R_o}{c}) +\frac{ \sin(\theta)}{2}\right]m}
\label{eq:RT},
\end{equation}
where $p(t,R_o) = e^{-j2\pi f_c (t-\frac{R_o}{c})}$. If the weight of the $m$th antenna is chosen as $w_m = e^{-jm\phi_{o}}$, (\ref{eq:RT}) can be written as
\begin{eqnarray}
r(t,R_o,\theta) &=& p(t,R_o)\sum_{m=0}^{M-1} 
e^{-j2\pi \left[f_o(t-\frac{R_o}{c}) +\frac{ \sin(\theta)}{2} + \frac{\phi_{o}}{2\pi}\right]m} \notag, \\
&=& p(t,R_o)\sum_{m=0}^{M-1} e^{-j m\pi \sin(\theta)}
  e^{-j2\pi [f_o(t-\frac{R_o}{c}) +\frac{ \phi_{o}}{2\pi}]m} \notag, \\
&=& p(t,R_o){\bm a }^{H}(\theta) {\bm s}(t;R_o) \notag,
\end{eqnarray}
where $\bm{ a}(\theta)=\left[ 1 \quad e^{j\pi \sin(\theta)} \quad  \cdots  \quad e^{j(M-1)\pi \sin(\theta)} \right]^{T}$ and $\bm{s}(t,R_o)=\left[ s_0(t-t_o)~~s_1(t-to)~\cdots~s_{M-1}(t-t_o) \right]^T$. Therefore, the average received power by the target located at range $R_o$ and in the direction $\theta$ can be written as 
\begin{align}
  P(\theta) &= \frac{1}{T} \int_{}^{}|r(t;R_o,\theta)|^{2}dt \notag \\
  &= \frac{1}{T}\int r(t;R_o,\theta) {r}^{H}(t;R_o,\theta)dt \notag \\
  &= \bm{a}^{H}(\theta) \bm{\Tilde{R}}\bm{a}(\theta), 
  \label{Power}
\end{align}
where
\begin{equation}
\Tilde{\bm{R}} = \frac{1}{T}\int{\bm{s}(t;R_o)\bm{s}^{H}(t;R_o)}dt,
\notag
\end{equation} 
is the correlation matrix of the transmitted waveforms. 
%
Using the timing diagram of individual pulses shown in Fig. \ref{fig2:accurate_time}, individual elements of the correlation matrix, $\bf{\tilde R}$, can be derived as
\begin{equation}
\Tilde{R}(m,n) = \frac{1}{T} \int_{t_o -(m -1)\tau(\theta)}^{t_o - (n -1)\tau(\theta) +T} s_{m}(t-t_o)s_{n}^{*}(t-t_o)dt. \label{CM1}
\end{equation}
By replacing the values of $s_m(t-t_o)$ and $s_n(t-t_o)$, (\ref{CM1}) can be  written as
\begin{align}
\Tilde{R}(m,n) = \frac{1}{T} \int_{t_o -(m -1)\tau(\theta)}^{t_o - (n -1)\tau(\theta) +T} e^{j2\pi [f_o(t-t_o) +\frac{ \phi_{o}}{2\pi}) ](n - m)}dt, \notag\\
= k (m,n) \frac{1}{T} \int_{t_o -(m -1)\tau(\theta)}^{t_o - (n -1)\tau(\theta) +T} e^{j2\pi \eta t} dt,~~\quad\quad\quad \notag \\
=  k(m,n) \frac{e^{j2\pi \eta t}}{ j2\pi \eta T}  \Biggr|_{t_o -(m -1)\tau(\theta)}^{t_o - (n -1)\tau(\theta) +T}, \qquad\qquad\qquad \notag
\end{align}
where $k (m,n) = e^{j2\pi \left[-f_o t_o +\frac{ \phi_{o}}{2\pi}\right](n - m)}$ and $\eta = f_o(n-m)$. %
Applying the limits we get
\begin{align}
\Tilde{R}(m,n) = k(m,n) \frac{e^{j2\pi \eta (t_o - (n -1)\tau(\theta) +T)}-e^{j2\pi \eta (t_o -(m -1)\tau(\theta))}}{ j2\pi \eta T}, \notag \\
= k (m,n) e^{j2\pi \eta (t_o - (n -1)\tau(\theta) +T)}
    \left(\frac{1-e^{j2\pi \eta ( (n -m)\tau(\theta) -T)}}{ j2\pi \eta T}\right)
    \label{B1},
\end{align}
Solving above equation further, we can write
\begin{eqnarray}
\tilde {R}(m,n) \!\!&=& \!\! -k(m,n) e^{j2\pi \eta (t_o - (n -1)\tau(\theta) +T)}e^{j\pi \eta ( (n -m)\tau(\theta) -T)}  \notag\\
&&\times \left(\frac{\sin(\pi \eta ( (n -m)\tau(\theta) -T))}{\pi \eta T}\right) \notag, \\
 \!\!&= \!\!&- k(m,n) e^{j\pi \eta (2t_o - (n + m - 2)\tau(\theta) + T)}  \notag\\
&&\times \left(\frac{\sin(\pi \eta ( (n -m)\tau(\theta) -T))}{\pi \eta T}\right) \label{eq:Rmn}.
\end{eqnarray}
In practice, the value of $\tau(\theta)\ll T$, therefore it can be ignored in (\ref{eq:Rmn}) and using this the average power in (\ref{Power}) can written as 
\begin{align}
P(\theta) &= \sum_{m=0}^{M-1}\sum_{n=0}^{N-1} e^{-jm\pi\sin(\theta)}\tilde{R}(m,n)e^{jn\pi \sin(\theta)}, \notag\\
&= \sum_{m=0}^{M-1}\sum_{n=0}^{N-1} e^{j(n-m)\pi \sin(\theta)}e^{j\pi(n-m)\left[f_oT+\frac{\phi_o}{\pi}\right]} \notag\\
&\quad\times \frac{\sin((n-m)\pi f_oT)}{(n-m)\pi f_oT}, \notag \\
&= \sum_{m=0}^{M-1}\sum_{n=0}^{N-1} e^{(n-m)\left[\pi \sin(\theta)+\pi f_oT + \phi_o \right]} \notag \\
&\quad\times \frac{\sin((n-m)\pi f_oT)}{(n-m)\pi f_oT}. \label{Power_sum1}
\end{align}
Assuming $f_{\theta}=\frac{\sin(\theta)}{2}$, (\ref{Power_sum1}) can be written as
\begin{align}
P(\theta) &= \sum_{m=0}^{M-1}\sum_{n=0}^{N-1} e^{j(n-m)\left[2\pi f_{\theta}+\pi f_oT + \phi_o \right]} \notag\\
&\quad\times \frac{\sin((n-m)\pi f_oT)}, \notag\\
& =  N + 2\sum_{n=1}^{N-1}(N-n)\frac{\sin(n\gamma)\cos{(n\kappa})}{n\gamma} 
\label{Power_sum4}
\end{align}
where $\kappa=(2\pi f_{\theta} +\pi f_oT +\phi_o)$ and $\gamma=\pi f_oT$. Using $\sin(\alpha)\cos{(\beta)}= \frac{1}{2}\left[ \sin(\alpha + \beta) +\sin(\alpha - \beta)\right]$ in  (\ref{Power_sum4}), it can be written as 
\begin{align}
      P(\theta) &= N + \sum_{n=1}^{N-1}\frac{1}{n\gamma}(N-n)\left[\sin(n(\gamma +\kappa)) + \sin(n(\gamma -\kappa)) \right]\notag \\
&= N +\frac{N}{\gamma}\underbrace{ \sum_{n=1}^{N-1}\left[\frac{\sin(n(\gamma +\kappa))} {n}+\frac{\sin(n(\gamma -\kappa))} {n}\right]}_{P_1(\theta)} \notag\\ 
&-\frac{1}{\gamma}\underbrace{\sum_{n=1}^{N-1}\left[\sin(n(\gamma +\kappa))+ \sin(n(\gamma -\kappa)) \right]}_{P_2(\theta)}
\label{eq:P1andP2}.
\end{align}
In (\ref{eq:P1andP2}), $P_1$ and $P_2$ can also be viewed as
\begin{align}
{P_1(\theta)} &= \underbrace{ \sum_{n=1}^{N-1}\left[\frac{\sin(n(\gamma +\kappa))} {n} \right] }_{P1_1(\theta)} + \underbrace{\sum_{n=1}^{N-1}\left[\frac{\sin(n(\gamma -\kappa))} {n}\right]}_{P1_2(\theta)}, \notag\\
{P_2(\theta)} &=  \underbrace{\sum_{n=1}^{N-1}\left[\sin(n(\gamma +\kappa))\right]}_{P2_1(\theta)}+\underbrace{\sum_{n=1}^{N-1}\left[ \sin(n(\gamma -\kappa)) \right]}_{P2_2(\theta)}
\label{eq:P2}.
\end{align}
The individual beampatterns of $P1$ and $P2$ along with their sub-terms $P1_{1}, P1_{2}$ and $P2_{1}, P2_{2}$ are respectively shown in Fig. \ref{fig:BW_Avg_Power1} and Fig. \ref{fig:BW_Avg_Power2}. While the overall beampattern of $ P$ is shown in Fig. \ref{fig:BW_Avg_Power}. 
Looking from the left side of Fig. \ref{fig:BW_Avg_Power1}, it can be observed, the location of the rising edges of $P1$ and $P1_1$ are almost the same. Let us assume, this location is denoted by $\theta_{1}$. Similarly, the location of the falling edges of $P1$ and $P1_2$ is almost the same. Assume this location is denoted by $\theta_{2}$. Similarly, the same relationship of $P2$ with its sub-terms can be observed in Fig. \ref{fig:BW_Avg_Power2}.
Therefore, the SE of FDA radar can be determined by finding the values of $\theta_1$ and $\theta_2$. It can be readily observed in (\ref{eq:P2}) that the zero crossings of $P1_1$ and $P1_2$ will respectively be the same as of $P2_1$ and $P2_2$. Therefore, we will only find the zero crossings of first pair, i.e. $P1_1$ and $P1_2$. Since $n\ne 0$, the term $P1_1$ will be zero iff
\begin{align}
    (\gamma +\kappa) &= \pm k\pi, \quad k=0,1,2,\cdots \notag\\
    (\pi f_oT +2\pi f_{\theta} + \phi_o + \pi f_oT) &= \pm k\pi. \qquad \qquad 
    \label{eq:+veZeroCrossing1}
\end{align}
The location of the first zero crossing of $P1_1$ can be found by inserting $k=0$ in (\ref{eq:+veZeroCrossing1})
\begin{align}
    (\pi f_oT +2\pi f_{\theta} +\pi f_oT + \phi_o) &= 0, \notag\\ 
    2\pi f_oT +2\pi f_{\theta} + \phi_o &= 0.   
    \label{eq:+veZeroCrossing2}
\end{align}
Solving (\ref{eq:+veZeroCrossing2}) yields the value of $\theta_1$ as
\begin{align}
    \theta_1 &= \sin^{-1}\left(-\left[ 2f_oT + \frac{\phi_o}{\pi}\right]\right) 
    \label{eq:thetaEnd}.
\end{align}
%
Similarly, the zero crossings of $P1_2$ can be derived by setting the argument of $\sin(\cdot)$ in the numerator of $P1_2$ equal to $k\pi$ as
\begin{eqnarray}
    n(\gamma -\kappa) &=& \pm k\pi, \quad k=0,1,2,\cdots
    \notag \\
    n\left(\pi f_oT -2\pi f_{\theta}-\pi f_oT - \phi_o\right) &=& \pm k\pi.
    \label{eq:-veZeroCrossing1}
\end{eqnarray}
The first zero crossing of the term $P1_2$ can be found by setting $k=0$ in (\ref{eq:-veZeroCrossing1}) as
\begin{align}
    (\pi f_oT -2\pi f_{\theta}-\pi f_oT - \phi_o) &= 0 \notag. 
\end{align}
The solution of the above equation yields the value of $\theta_{2}$
\begin{equation}
    \theta_{2} = \sin^{-1}\left({- \frac{\phi_o}{\pi}}\right)
    \label{eq:thetaStart}.
\end{equation}
Using (\ref{eq:thetaEnd}) and (\ref{eq:thetaStart}), SE of FDA radar can be found as
\begin{align}
  \textrm{SE} &= \theta_{2} - \theta_{1}, \notag\\ 
              &= \sin^{-1}\left({- \frac{\phi_o}{\pi}}\right) -\sin^{-1}\left(-\left[ 2f_oT  + \frac{\phi_o}{\pi}\right]\right).
     \label{eq:Beamwidth}
\end{align}
Expanding sine function in (\ref{eq:Beamwidth}) using the Maclaurin series again as described in Sec. \ref{subSec:BeamwidthDerivation}, the approximate value of SE can be easily derived as   
\begin{align}
\textrm{SE} \approx 2f_o T +  \frac{2\phi_o}{\pi} (f_o T)^2. \label{eq:Beamwidth2}
\end{align}
Observing the above equation, it can be said that SE depends mainly on the product of FO and pulse duration i.e. $f_{o}T$. The effect of $\phi_o$ on the SE will be minimal. For further understanding of the SE, instantaneous FDA radar beampatterns at different time instances are shown along with the average beampattern in Fig. \ref{fig:Power,Instant_Avg}.  

%
\begin{figure}
    \centering
    \captionsetup{justification=centering}
    \includegraphics[scale=0.6]{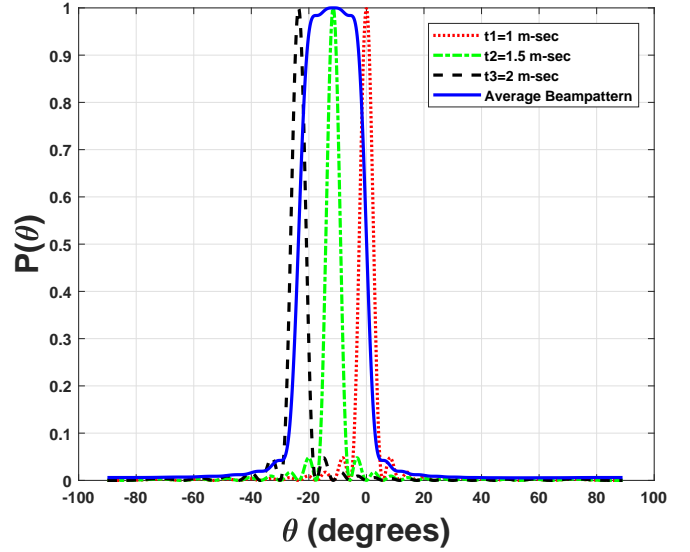}
    \caption{Instantaneous and average beampatterns. Initial phase $\phi_o=0^o$, $M = 20$,$f_o=200$Hz, $T = 1m$s.}
    \label{fig:Power,Instant_Avg}
\end{figure}

%
%
%
%
\subsection{Bound on the Selection of \enquote*{$f_oT$}}
As mentioned in the last section the SE depends on the value of $f_oT$. If $\phi_o=0$, the value of $f_oT$ can be found using  (\ref{eq:+veZeroCrossing2}) as
\begin{align}
    2\pi f_oT +2\pi f_{\theta} &= 0 
    \label{eq:foT}.
\end{align}
Solving (\ref{eq:foT}) yields
\begin{align}
    f_oT   &= - f_{\theta} \notag \\
           &= -\frac{\sin(\theta)}{2} \le 0.5 
    \label{eq:foT1}.
\end{align}
Therefore, (\ref{eq:foT1}) gives us the criteria to select the value of $f_oT$.  

As we have seen the beamwidth of conventional FDA radar depends inversely on the number of antennas while its SE depends on the product of $f_o$ and $T$. Most of the reported algorithms in the literature produce narrow instantaneous beampatterns that has short dwell time and becomes further narrow as the number of antennas is increased. To increase the dwell time the instantaneous beampatterns should be wide. To achieve the wide beampatterns, weights of the FDA radar can be optimized. The proposed algorithms in the literature use computationally complex algorithms, such as convex optimization, to find such weights. To reduce the computational complexity to optimize the weights, in the following section we exploit DFT and propose a closed-form solution.    
\section{Proposed Transmit beampattern Design} {\label{Sec:PrpsdSchm}}
In FDA radar FO is kept much lower than the carrier frequency or $f_o \ll f_c$. As a consequence, the factor $\frac{m^2 f_o d \sin{\theta}}{c} \ll  \frac{m f_c d \sin{\theta}}{c}$ and it can be ignored in (\ref{eq:AF1}) to write the equation in more compact form as
\begin{align}
\AF(t,R_o,\theta) &= \sum_{m=0}^{M-1} w_{m}
  e^{-j2\pi [mf_o(t-\frac{R_o}{c}) +m\frac{ \sin(\theta)}{2}) ]} \notag\\
   &=  \sum_{m=0}^{M-1} w_{m}
  e^{-j2\pi f_{\theta}(t) m}   \label{eq3:BP_cr1},
\end{align}
where $f_{\theta}(t)=f_o(t-\frac{R_o}{c}) + \frac{ \sin(\theta)}{2}$. It should be noted that at $t=\frac{R_o}{c}$, the value of $f_{\theta}(t) = \frac{\sin(\theta)}{2}$ and it will lie between $-0.5$  and $0.5$.  Therefore, AF can be written as
\begin{align}
\AF{\left(t=\frac{R_o}{c},R_o,\theta \right)} &= \sum_{m=0}^{M-1} w_{m}
e^{-j2\pi f_{\theta} m} \label{eq3:BP_cr2},
\end{align}
which can be considered as the equation of discrete-Fourier-transform (DFT) \cite{Book:SNS}. Therefore, it can be said that at $t=\frac{R_o}{c}$, the AF of FDA radar is the DFT of weights $w_m$. Conversely, if the values of $\AF\left(t=\frac{R_o}{c};R_o,\theta \right)$ with respect to $\theta$ are known, using the inverse-DFT (IDFT) the corresponding weights can be easily calculated as 
\begin{equation}
w_{m}  = \sum_{f_\theta=-0.5}^{0.5}\AF\left(t=\frac{R_o}{c},R_o,\theta \right)
e^{j2\pi f_{\theta} m}. \label{eq:IDFT}
\end{equation}
If AF of FDA radar is know the corresponding beampattern can be found by just taking the square of its absolute value as given in (\ref{eq:BP}). Therefore, (\ref{eq:IDFT}) can be exploited to find the weights of FDA radar for the desired beampattern. 

To explain the process of finding the weights for the desired beampattern, consider $B(t,R_o,\theta)$ represents the desired beampattern. The theoretical spatial region that a ULA radar can illuminate is a bounded region between -90 degrees to +90 degrees defined as $\theta\in \{-90, 90\}$, which corresponds to $f_\theta \in \{-0.5, 0.5\}$. To focus the transmitted power within a given region at a distance $R_o$, the overall region can be divided into a number of grid points, with each grid point representing an angular location in the region. To illuminate an angular location assign one to a corresponding grid point otherwise assign zero to it. To achieve the desired beampattern, calculate the corresponding weights using (\ref{eq:IDFT}).

To explain the working and asses the performance of proposed scheme two numerical examples are given in the following. 

\begin{figure}[h]
    \centering
    \centerline{\psfig{figure=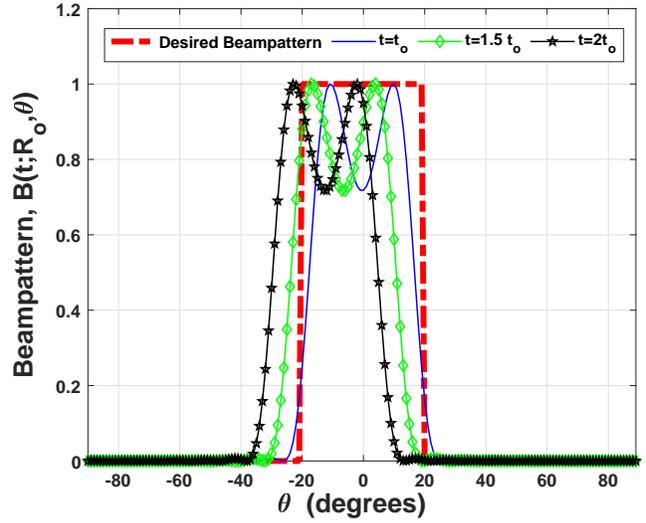,width=3.70in} }
    \caption{Desired and designed single beampattern for $R_o=300k$m, $M=20$, $f_o=100 $Hz, $T = 1 m$s.} \label{Fig:Desired_Designed_SBP}
\end{figure}

\begin{figure}[h]
    \centering
    \centerline{\psfig{figure=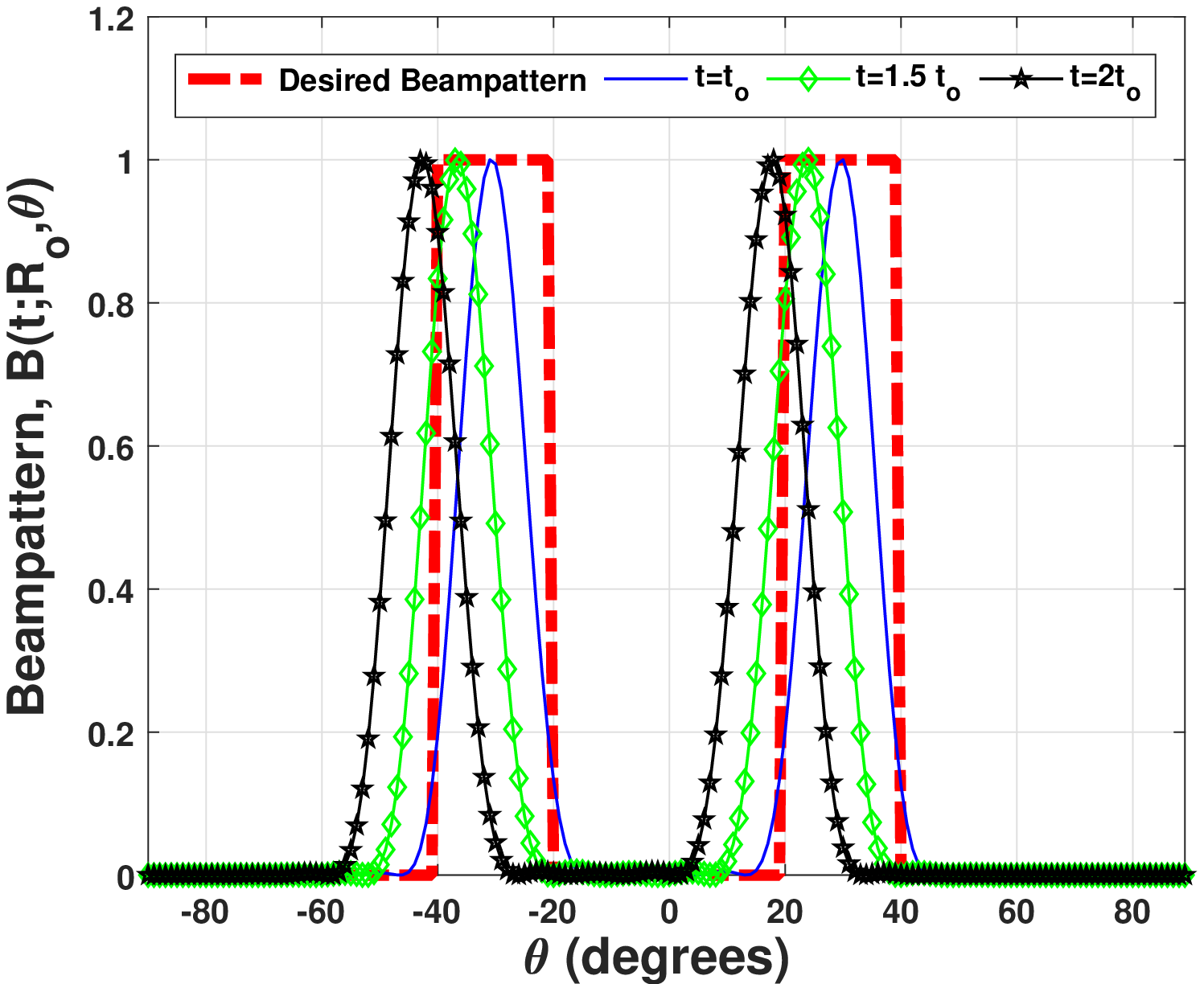,width=3.70in} }
    \caption{Desired and designed beampattern for $R_o=300k$m, $M=20$, $f_o=100 $Hz, $T = 1 m$s.} \label{Fig:Desired_Designed_DBP}
\end{figure}

In the first example, the desired region is defined by $\Theta_1 \in \left[-20, 20\right]$ while in the second example the desired region is defined by $\Theta_2 \in \left[-20, -40\right]U\left[20, 40\right]$.  For both examples the number of transmit antennas used is $M=20$ and the target is at $R_o=300k$m. The desired and designed beampatterns at different instances of time $t=\frac{R_o}{c}$, $t=\frac{1.5R_o}{c}$, and $t=\frac{2R_o}{c}$ are shown in Fig. \ref{Fig:Desired_Designed_SBP} and \ref{Fig:Desired_Designed_DBP}. As can be seen at $t=\frac{R_o}{c}$, the designed beampattern follow the pattern of desired beampattern. By increasing the number of antenna, the pattern of the designed and desired beampatterns can be further matched. 
It should be noted, as shown in the figure, for $t>\frac{R_o}{c}$ the beampattern is shifting towards left. The reason for this is straightforward and can be noticed by rewriting (\ref{eq3:BP_cr1}) as
\begin{align}
\AF(t;R_o,\theta) &= \sum_{m=0}^{M-1}  e^{-j2\pi f_o\left(t-\frac{R_o}{c}\right)m}
  w_{m} e^{-j2\pi f_\theta m} \label{eq3:BP_cr1a}.
\end{align}
Looking (\ref{eq3:BP_cr1a}) in terms of index $m$ and $f_\theta$ domain, it can be observed that the modulation theorem given by \cite{Book:SNS}
\begin{equation}
X(k + k_o) \xLongleftrightarrow{\textrm{  DFT  }} e^{-j2\pi k_{o}t}x(t) \notag
\end{equation}
can be applied to find the beampattern at any arbitrary value of $t>t_o$. According to this theorem, the shift in the beampattern at any value of time $t>t_o$ will be  $-f_o\left(t-\frac{R_o}{c}\right)$.
%
%
%
%
It can also be noticed in the figures, the designed  beampatterns are slightly moving their spatial focus with respect time, which is inherited problem of FDA. It is well known that the dwell time of conventional FDA radar system is so less that it may miss the weak targets with high probability. As it can be observed in the figures, the dwell time of proposed scheme is dependent on the spatial region chosen. The higher the desired spatial region the higher will be the dwell time. So the proposed DFT-FDA scheme offers the control over the dwell time. From this control ability the DFT-FDA has the advantage to address the problem of shorter dwell time of conventional FDA.   
%
%
%
%
\section{Simulation Results}{\label{Sec:Simulations}}
In this section, to compare the performance of the proposed algorithm with the existing algorithms a number of simulations are performed. In all of the following simulations, a ULA FDA radar is considered, the number of antennas $M=20$, the carrier frequency $f_c=5$GHz, the FO $f_o=100$Hz, the initial phase of the signal $\phi_o=0$ degrees, and the duration of the transmitted pulse $T=1m$s. In the following  simulations, the beampattern with respect to range are discussed, therefore the contribution of path-differences due to the antennas elements will be ignored.    

\begin{figure*}[htbp!]
    \centering
    \begin{subfigure}{.3\linewidth} 
        \includegraphics[scale=0.4]{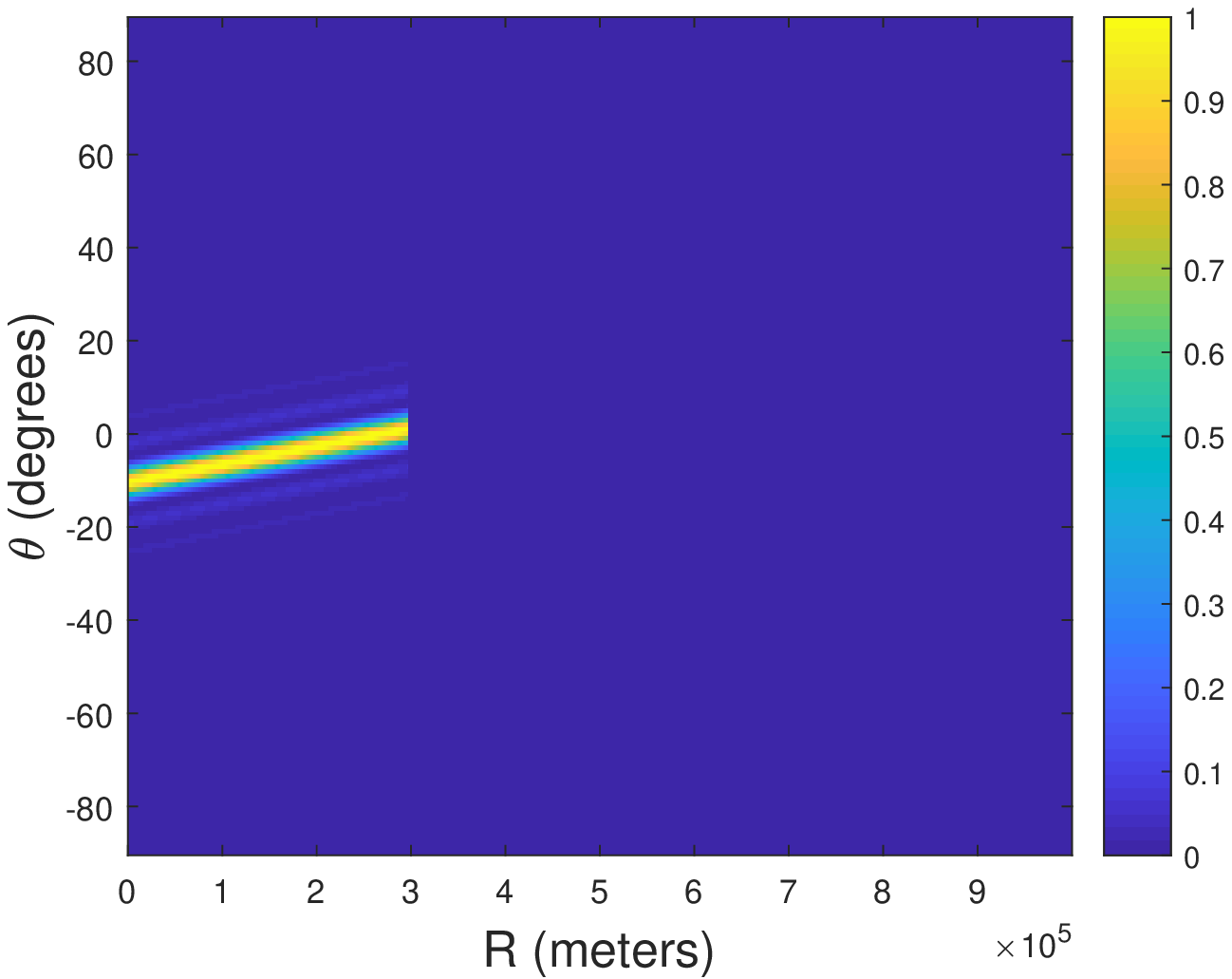}
        \caption{}
    \end{subfigure}
    \hskip0.5em
    \begin{subfigure}{.3\linewidth}
        \includegraphics[scale=0.4]{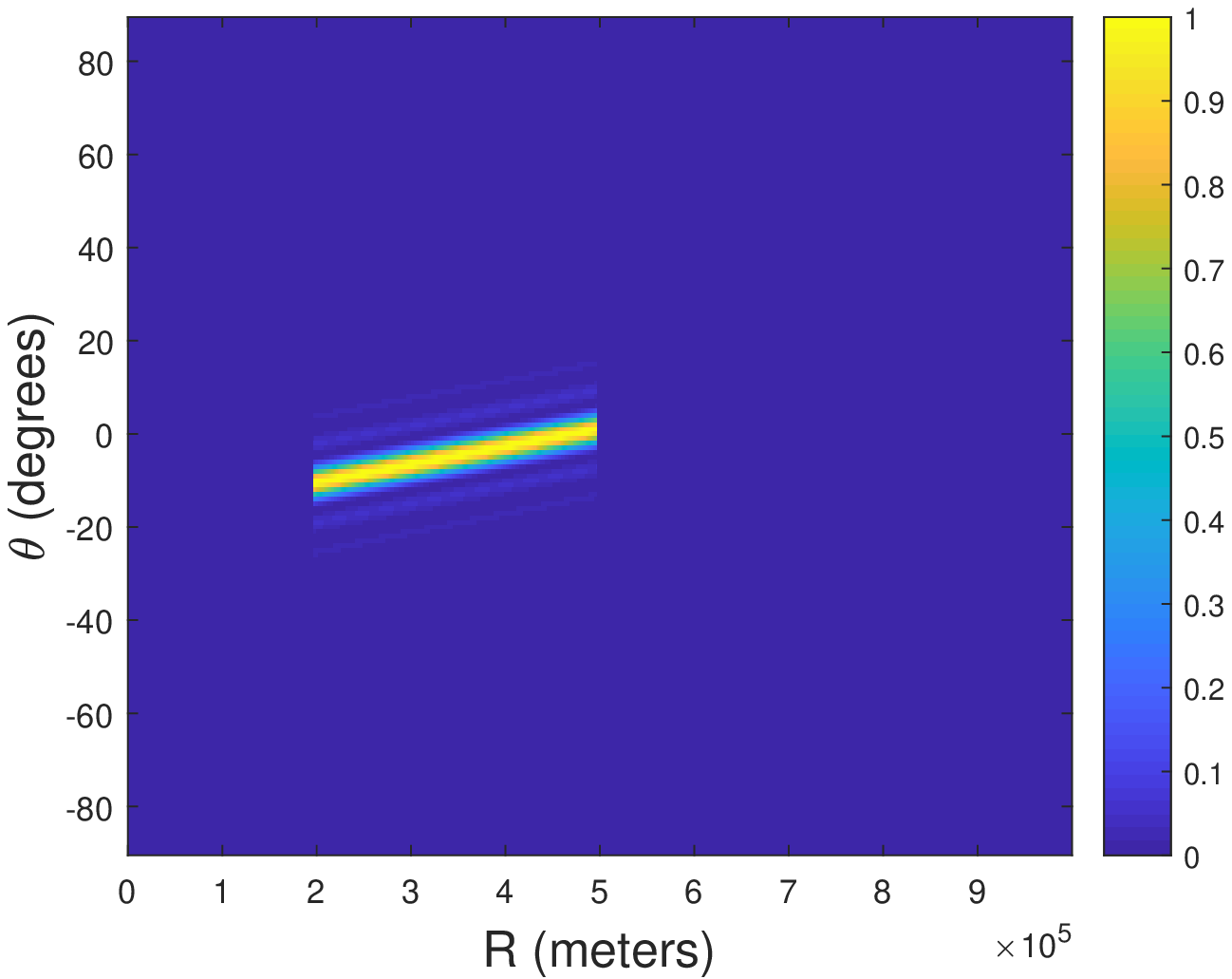}
        \caption{}
    \end{subfigure}
    \begin{subfigure}{.3\linewidth}
        \includegraphics[scale=0.4]{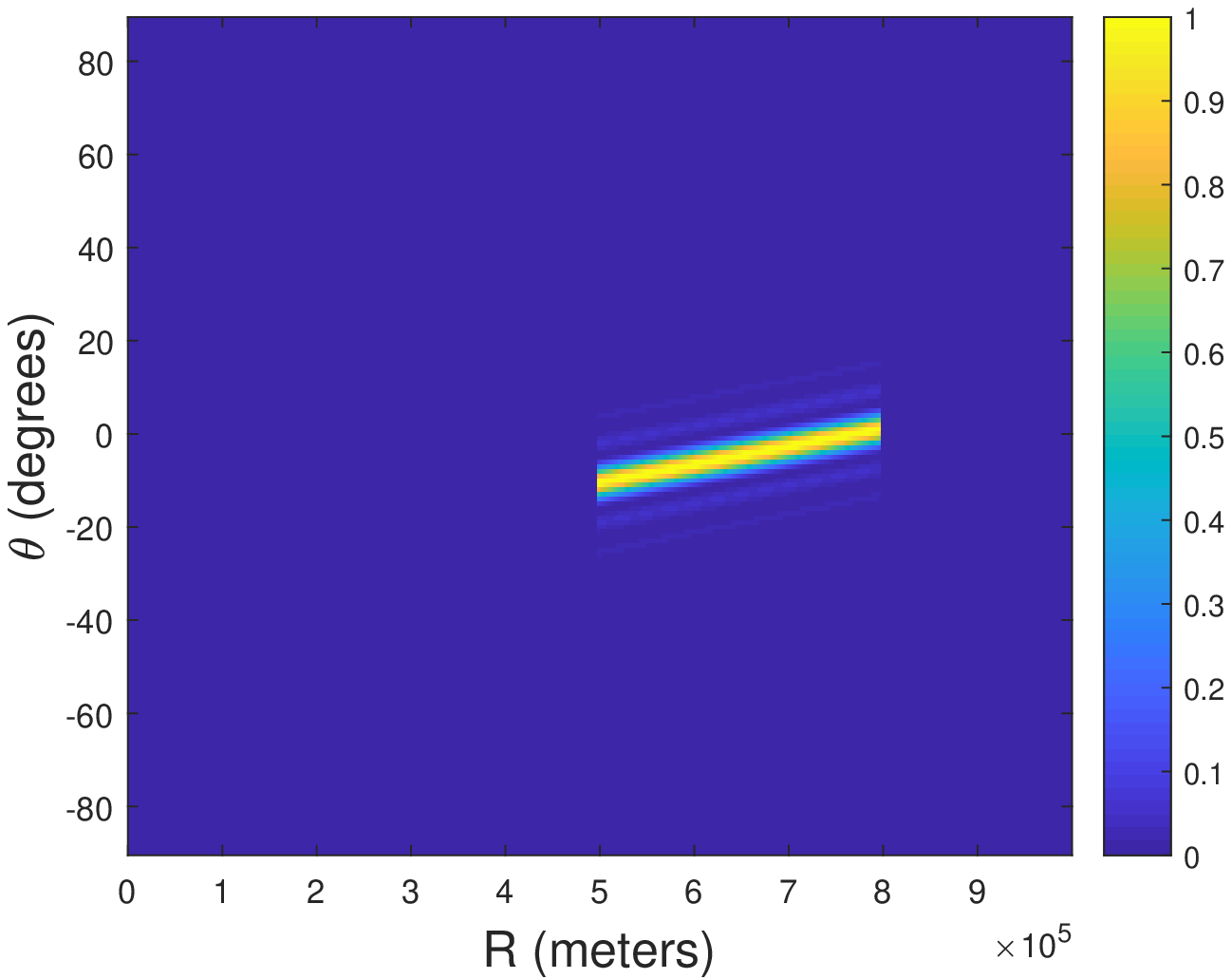}
        \caption{}
    \end{subfigure}
    \caption{conventional FDA radar beampattern at (a) $t_o= 1m$s (b) $t_o= 1.66m$s (c) $t_o= 2.66m$s.}
    \label{Simulation:FDA_Standard}
\end{figure*}

In the first simulation, conventional FDA radar with $w_m=1$ for $m=1,2,\ldots, M-1$ is considered \cite{Timeinvariant6chen}. The corresponding beampatterns with respect to range at different time instances are shown in Fig.\ref{Simulation:FDA_Standard}. Since the duration of pulse $T=1m$s, examining (\ref{eq:AF6}) it can be noticed that the beampattern will be symmetric when
$\frac{R_o}{c}=1m$s or $R_o=300k$m. At $t=1m$s the leading edge of the transmitted pulse from the reference antenna  will be at $R_o=300k$m and the trailing edge will be at $R_o=0$. In this case, targets at ranges more than $300k$m will not be illuminated. On the other hand all targets at ranges  less than $300k$m will be illuminated by all transmitted signals. For these ranges, in the AF expression (\ref{eq:AF6}), the term $f_o\left(t-\frac{R_o}{c}\right)$ will be positive and as a consequence the symmetric beampattern will be shifted towards the negative spatial angle. This effect can be noticed in Fig. \ref{Simulation:FDA_Standard}.a. 
At $t=1.66m$s the beampattern will be symmetric at range $R_o=500k$m. Similar to previous case at ranges more than $500k$m none of the target will be illuminated, at ranges less than 500$k$m all targets will be illuminated, and beampattern will be shifted towards the negative spatial angle corresponding to the ranges less than 500$k$m. These effects can be seen in Fig. \ref{Simulation:FDA_Standard}.b. 
Finally, same effects can be observed in Fig. \ref{Simulation:FDA_Standard}.c at $t=2.66m$s. 

\begin{figure*}[htbp!]
    \centering
    \begin{subfigure}{.3\linewidth}
        \includegraphics[scale=0.35]{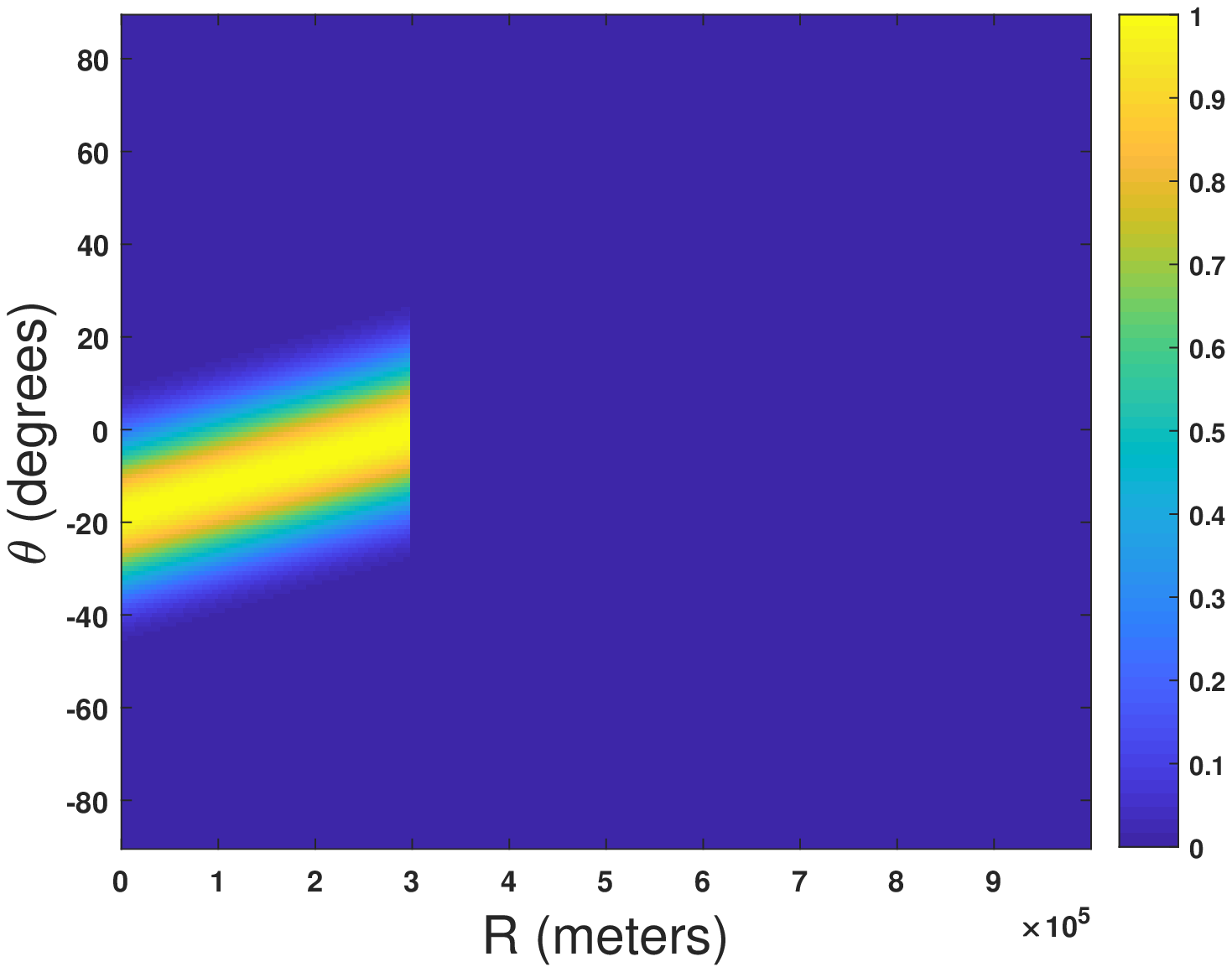}
        \caption{}
    \end{subfigure}
    \hskip0.5em
    \begin{subfigure}{.3\linewidth}
        \includegraphics[scale=0.35]{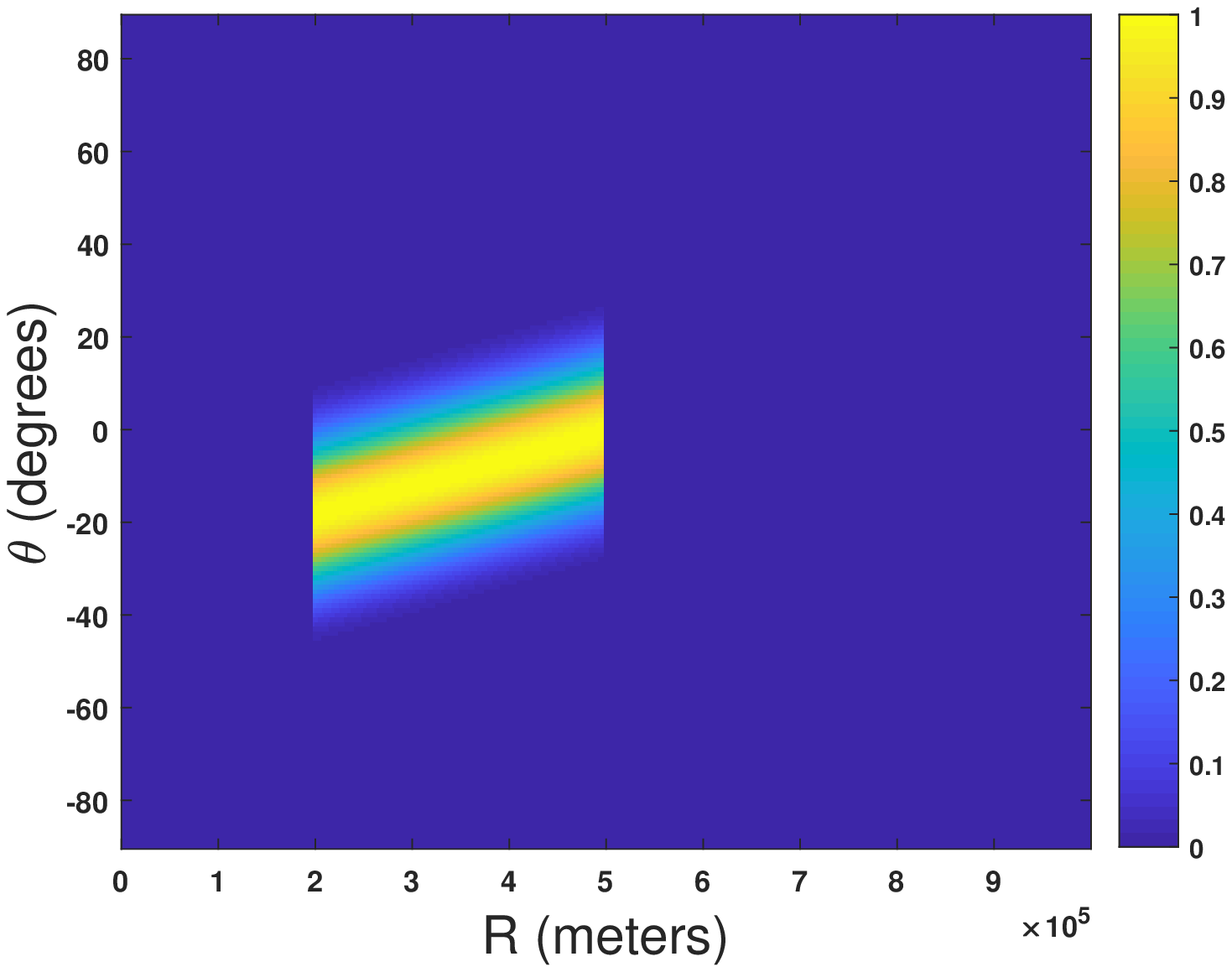}
        \caption{}
    \end{subfigure}
    \begin{subfigure}{.3\linewidth}
        \includegraphics[scale=0.35]{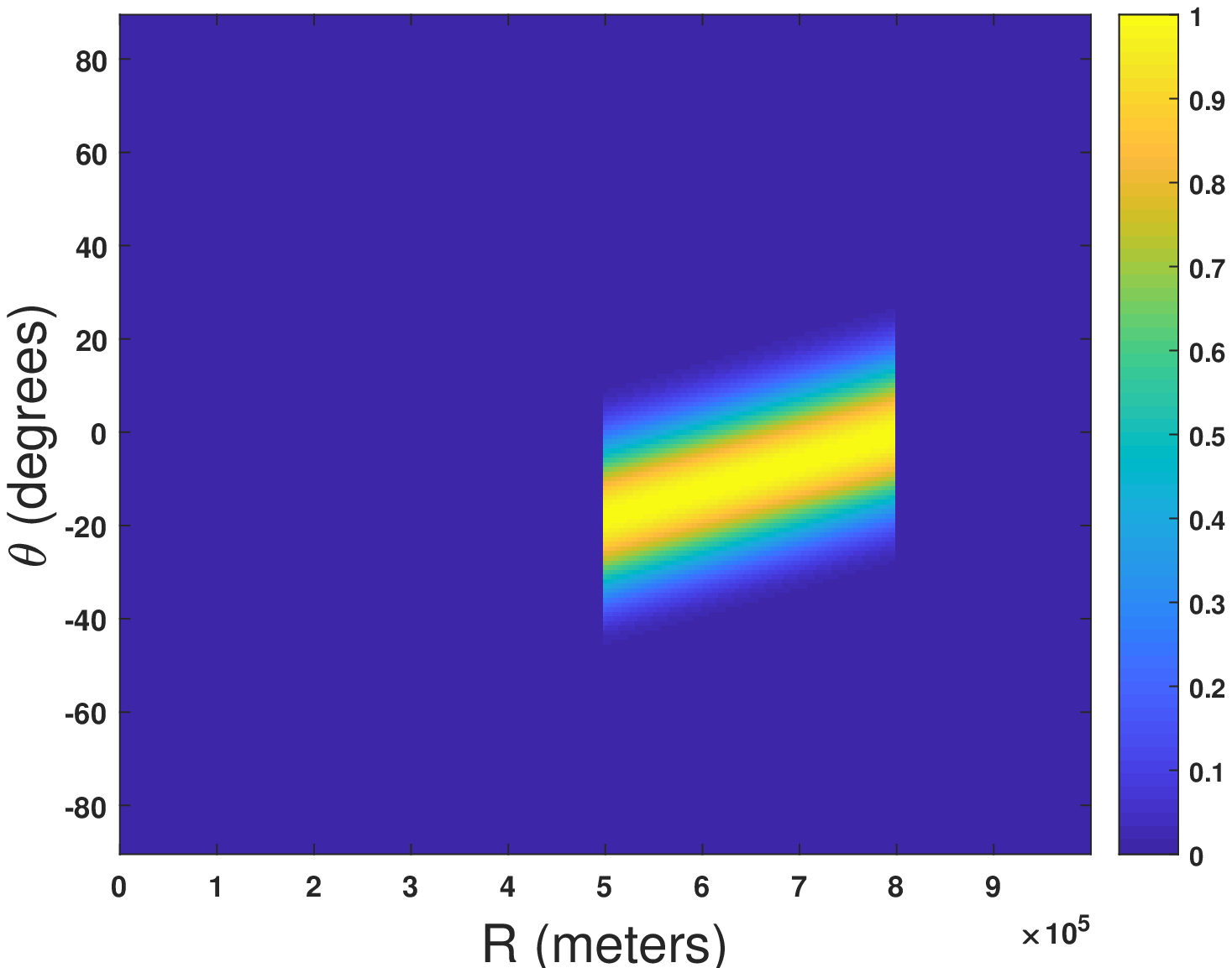}
        \caption{}
    \end{subfigure}
    \caption{Desired single main-lobe proposed FDA radar beampattern at (a) $t_o= 1m$s (b) $to= 1.66m$s (c) $t_o= 2.66m$s.}
    \label{Simulation:FDA_Proposed1}
\end{figure*}

To increase the dwell time, conventional FDA radar cannot be used to design wide beampatterns. The proposed algorithm can be exploited for this objective. Therefore, in the second simulation, to focus the transmitted power in the wider region defined by $\Theta_1 \in \left[-20,20\right]$ at $R_o=300k$m the proposed algorithm is used.  Simulation results at different time instances are shown in Fig.\ref{Simulation:FDA_Proposed1}.a-Fig.\ref{Simulation:FDA_Proposed1}.c. Wide beampatterns for longer dwell times along with the angular shifts in beampattern for ranges corresponding to selected times similar to previous case can be observed in these figures. 

\begin{figure*}[htbp!]
    \centering
    \begin{subfigure}{.3\linewidth}
        \includegraphics[scale=0.4]{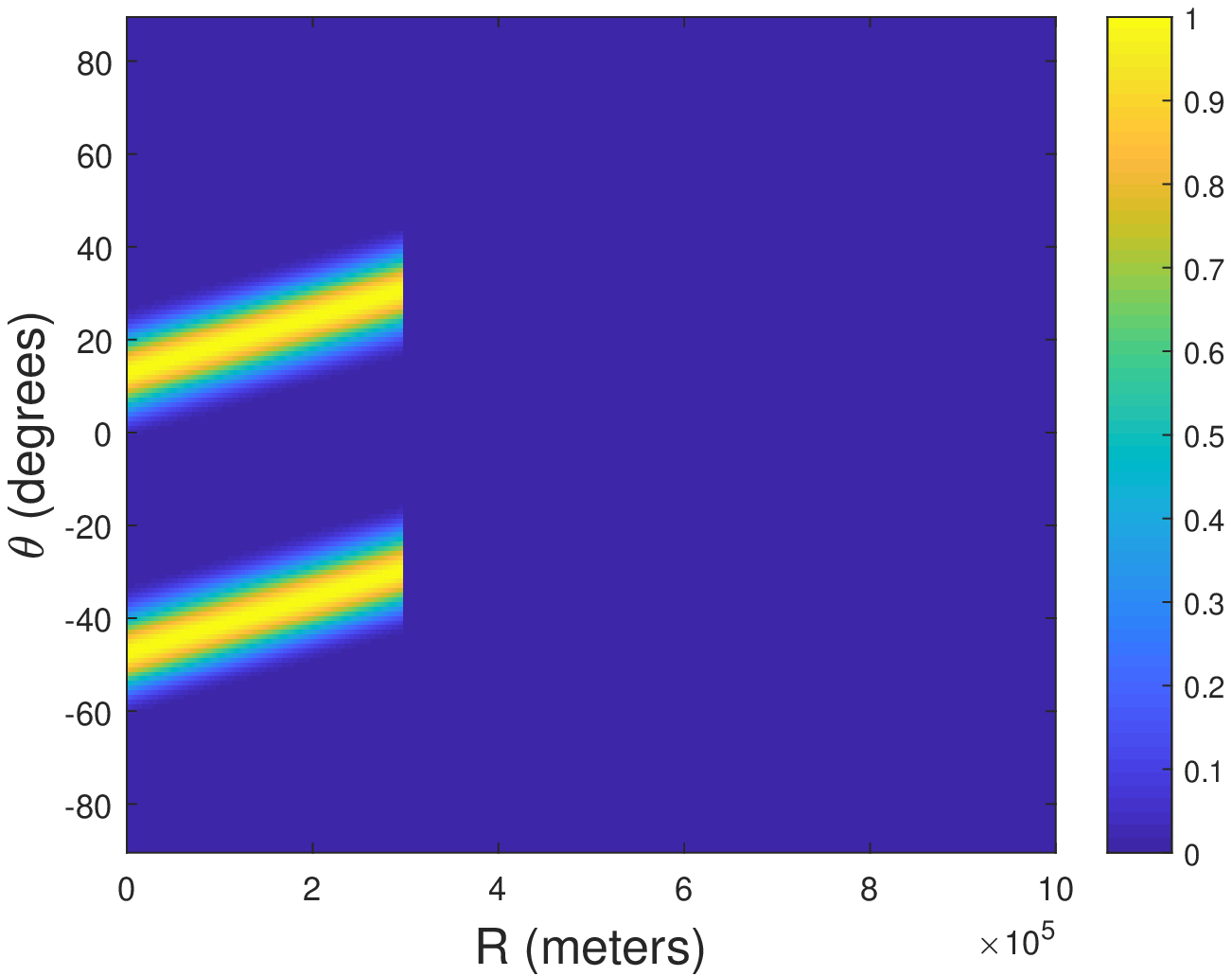}
        \caption{}
    \end{subfigure}
    \hskip0.5em
    \begin{subfigure}{.3\linewidth}
        \includegraphics[scale=0.4]{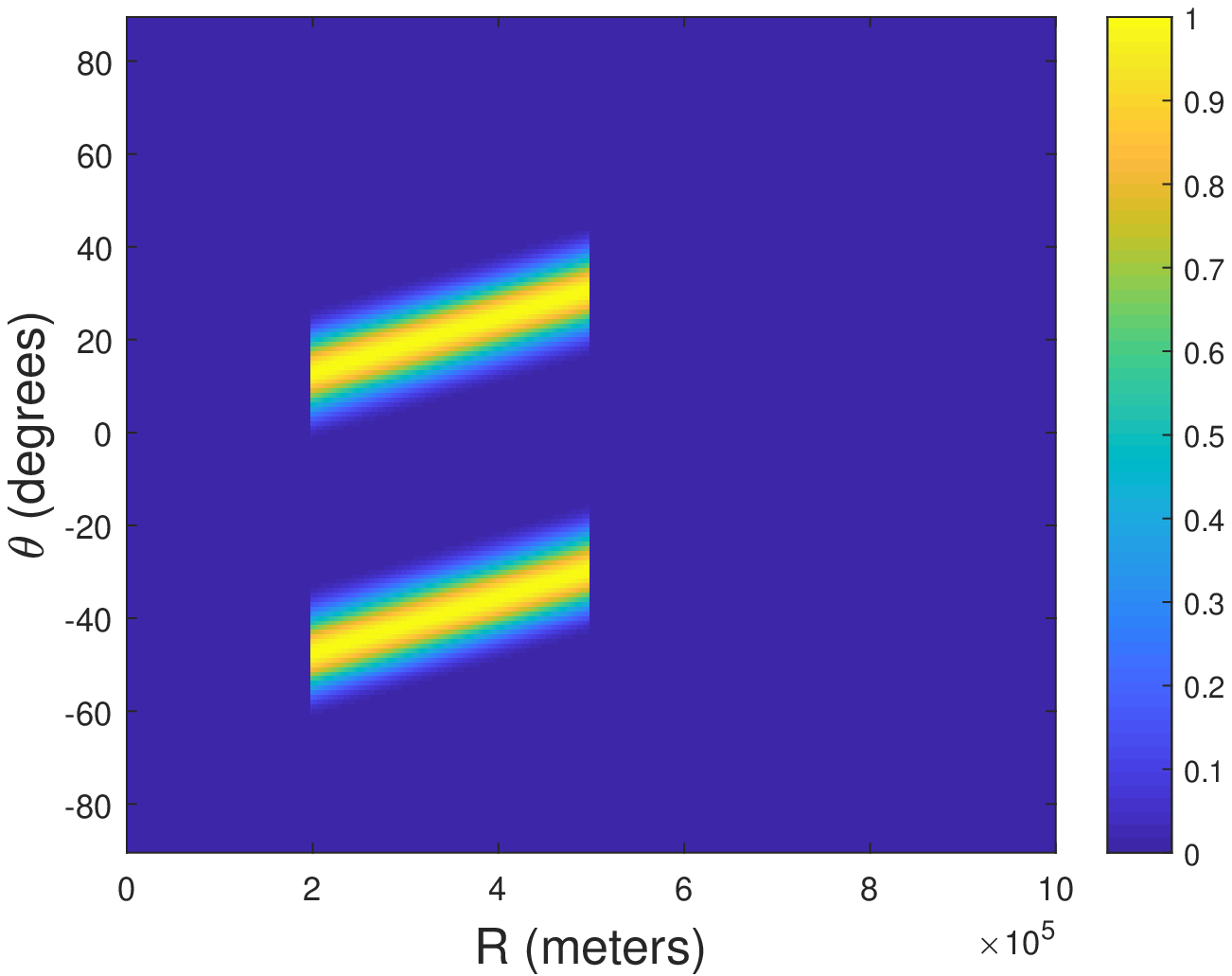}
        \caption{}
    \end{subfigure}
    \begin{subfigure}{.3\linewidth}
        \includegraphics[scale=0.4]{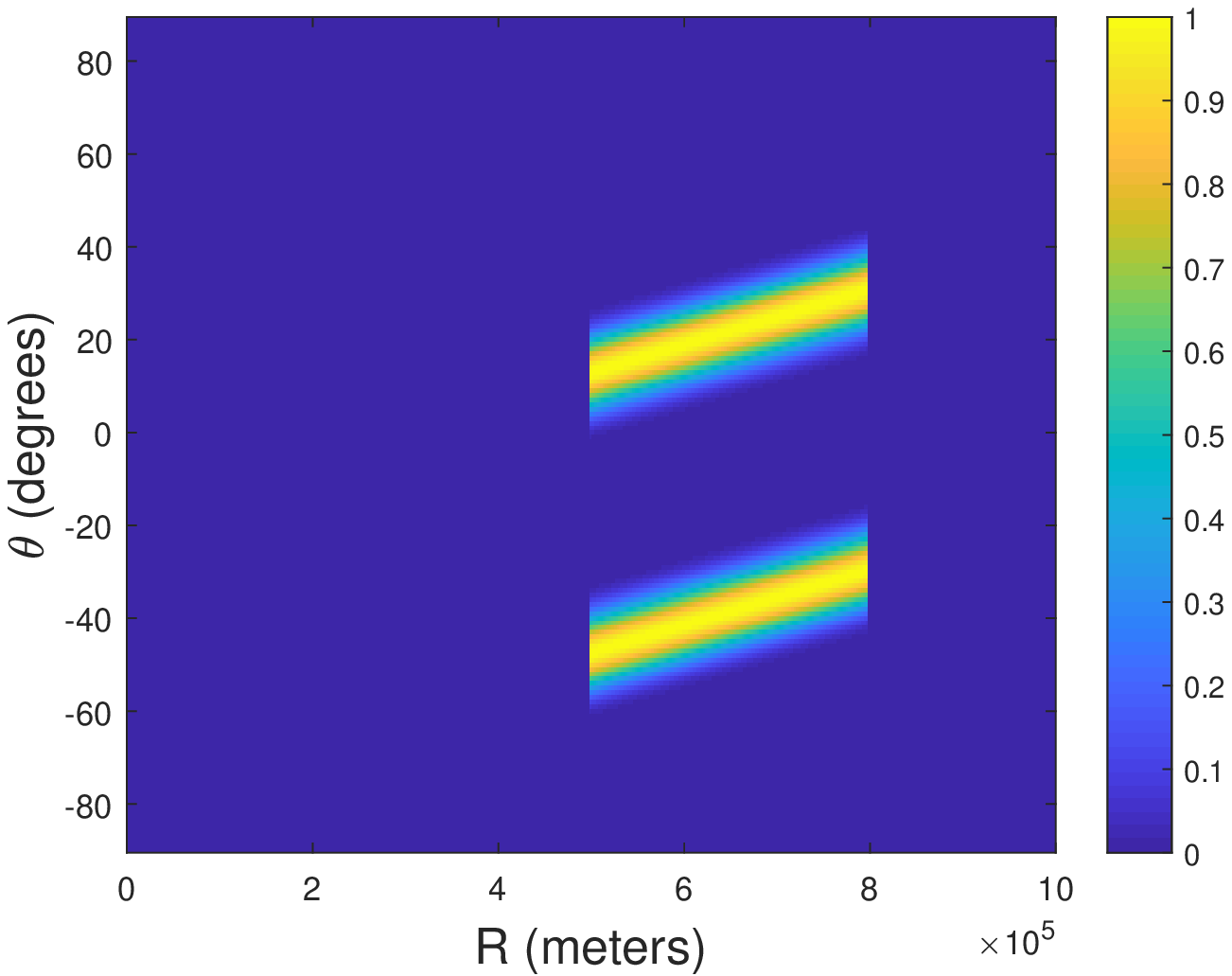}
        \caption{}
    \end{subfigure}
    \caption{Desired two main-lobe proposed FDA radar beampattern at (a) $t_o= 1m$s (b) $to= 1.66m$s (c) $t_o= 2.66m$s.}
    \label{Simulation:FDA_Proposed2}
\end{figure*}

Similarly, if we want to illuminate two regions at the given range the proposed algorithm can also be used. 
Therefore, in the third simulation, to focus the transmitted power between two regions defined by $\Theta_2 \in \left[-20, -40\right]U\left[20, 40\right]$ at a range $R_o=300k$m the proposed algorithm is used. Corresponding beampatterns at different time instances can be seen in Fig.\ref{Simulation:FDA_Proposed2}.a-Fig.\ref{Simulation:FDA_Proposed2}.c. Similar to the results presented in the previous two simulations, shifts in the beampatterns can be observed for ranges less than selected $R_o$. 

\begin{figure*}[htbp!]
    \centering
    \begin{subfigure}{.3\linewidth}
        \includegraphics[scale=0.4]{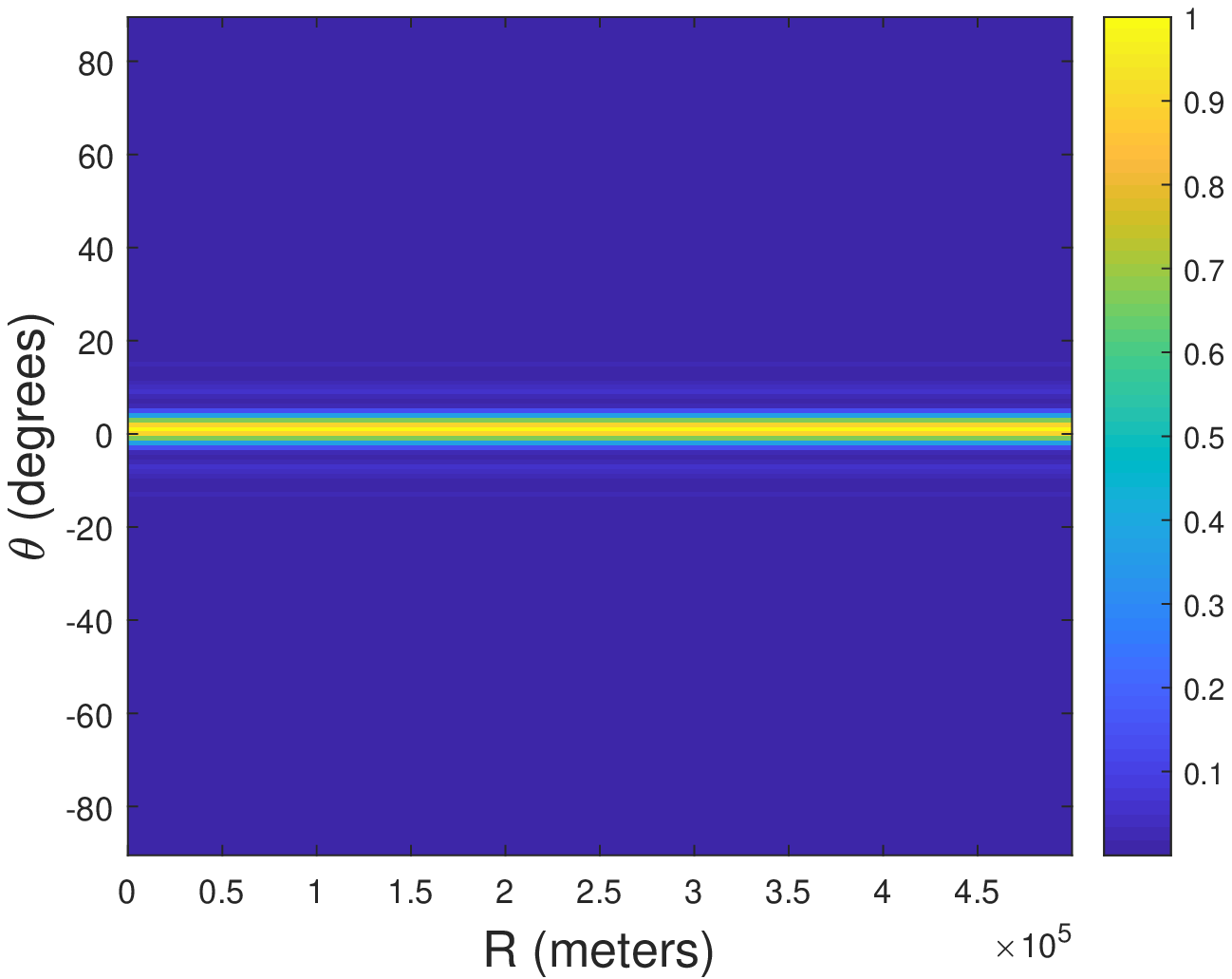}
        \caption{}
    \end{subfigure}
    \hskip0.5em
    \begin{subfigure}{.3\linewidth}
        \includegraphics[scale=0.4]{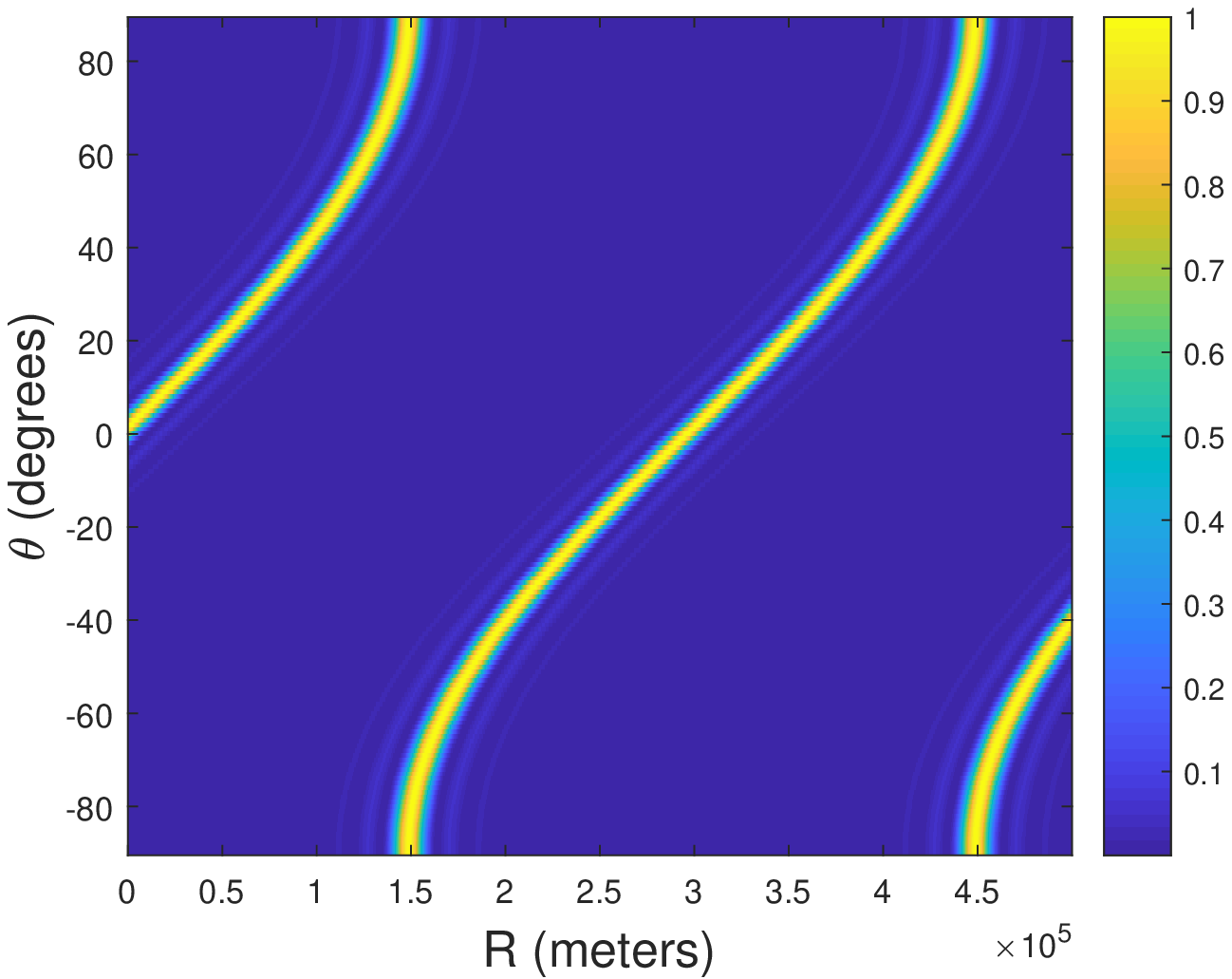}
        \caption{}
    \end{subfigure}
    \begin{subfigure}{.3\linewidth}
        \includegraphics[scale=0.35]{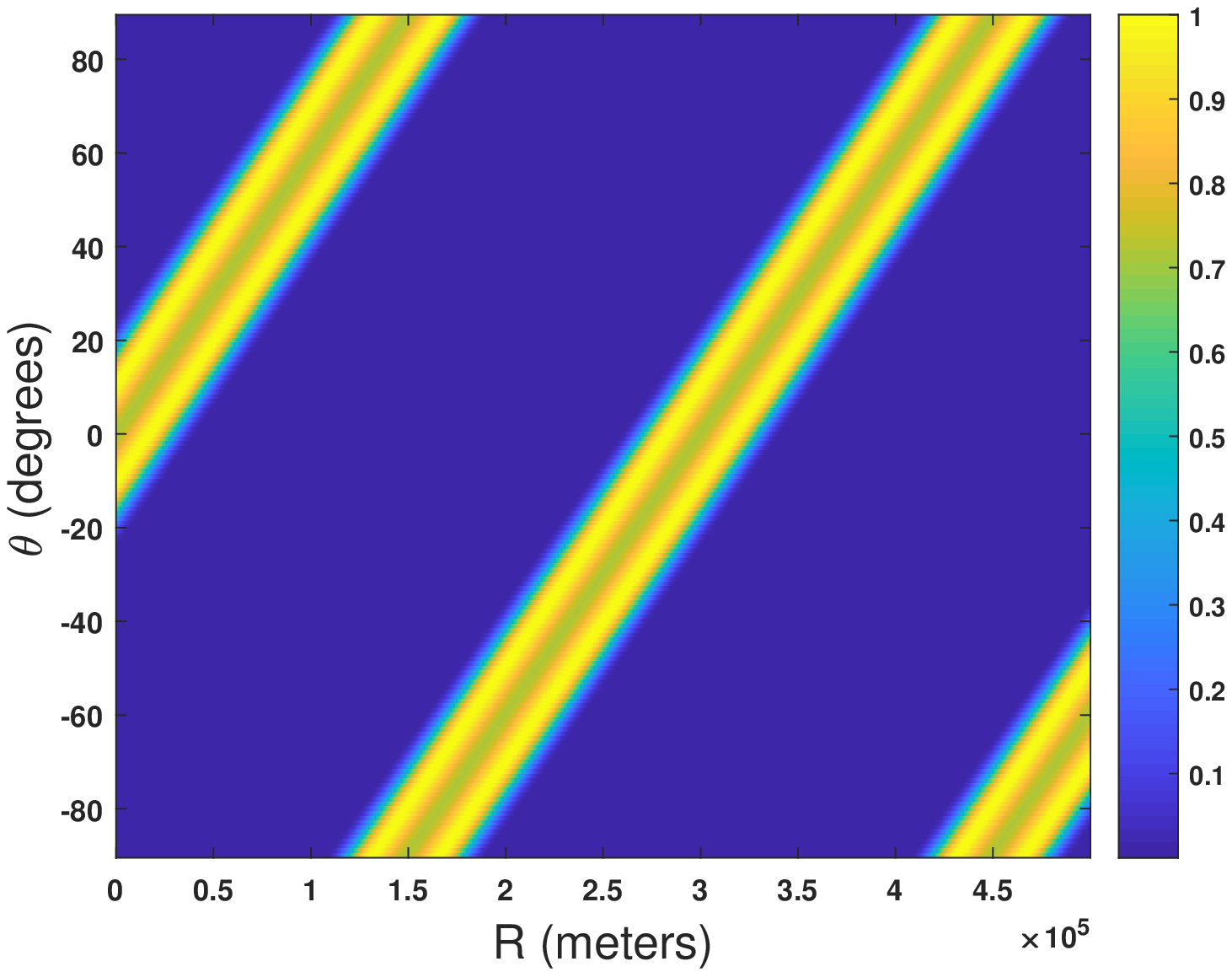}
        \caption{}
    \end{subfigure}
    \caption{Comparison of different schemes beampatterns  (a) PAR (b) FDA radar (c) Proposed DFT based FDA radar}.
    \label{Simulation:Phased_FDA_Proposed}
\end{figure*}

In final simulation, the beampatterns of different schemes, where FOs are  continuously (continuous-wave FDA radar) applied instead of short duration (pulsed FDA radar) are compared in Fig. \ref{Simulation:Phased_FDA_Proposed}. Here for all schemes, FO $f_o=1k$Hz and rest of the parameters are same as in the above simulations. The beampattern of PAR is shown in  Fig.\ref{Simulation:Phased_FDA_Proposed}.a. It can be noticed here that the beampattern remains same for all ranges and does not tilt. The beampattern of continuous-wave FDA radar is shown in Fig.\ref{Simulation:Phased_FDA_Proposed}.b. Here, a non-linear \enquote{$S$} shaped tilt in the beampattern with respect to range can be observed. Finally, the desired width beampattern of the proposed scheme for increased dwell time is shown in Fig.\ref{Simulation:Phased_FDA_Proposed}.c. Here, although the tilt of beampattern is a function of range but it is smooth and linear compared to the continuous-wave FDA radar.

It should also be noted that the tilt and periodicity of conventional FDA radar beampattern are respectively the function of $\sin^{-1}(2f_oT)$ and $\frac{c}{f_o}$. The difference in the tilt of beampattern in the first (Fig. \ref{Simulation:FDA_Standard}) and final (Fig.  \ref{Simulation:Phased_FDA_Proposed}) simulations respectively of conventional pulsed and continuous FDA radar is due to the difference in the FOs. In the first simulation $f_o=100$HZ while in the final simulation $f_o=1k$HZ.
\section{Conclusion and Future Work}{\label{sec:con}}
In this paper, we have proposed a novel closed-form solution to optimize the weights of FDA radar for the desired beampattern. The proposed solution exploits the DFT to synthesis the transmit beampattern for the desired spatial regions. The proposed solution is simple, flexible and more tractable and it may spawn multiple applications of FDA radar that are not accessible by using conventional FDA and phased array radar. We have also reanalyzed Array factor and beampattern expression for FDA radar and devised the correct time-range constraint that must be incorporated for more realistic and more effective response. In addition to that an average beampattern expression is calculated to calculate SE and relation between FOs and pulse duration. By reanalyzing  with the correct signal model, we have verified and concluded that it is impossible to design a beampattern that will remain fix and illuminate the target for whole duration of pulse, however dwell time ca be increased using proposed scheme. The further potential of this idea will be exploited in Receiver design and in planar array FDA radar system. 

 \bibliographystyle{ieeetr}
 \bibliography{JournalRadar}
\end{document}